\documentclass[UTF8]{cpbtex}
\usepackage{CJKutf8}
\usepackage{bbding}
\usepackage{pifont}
\usepackage{soul}
\usepackage{mathrsfs}
\usepackage{enumitem}
\usepackage{hyperref}
\usepackage{setspace}

\usepackage{graphicx}
\usepackage{physics}
\usepackage{bm,amsmath,amssymb,accents}

\usepackage{dcolumn}
\usepackage{xcolor}

\usepackage{bm}

\usepackage{algorithm}
\usepackage{algorithmicx}
\usepackage{algpseudocode}
\usepackage{placeins} 
\usepackage{booktabs} 
\makeatletter

\newcommand{\Rmnum}[1]{\expandafter\@slowromancap\romannumeral #1@}
\makeatother

\begin{document}

\title{Emergent topological ordered phase   for the Ising-XY Model revealed by cluster-updating Monte-Carlo method}


\author{Heyang Ma(\begin{CJK}{UTF8}{gbsn}
马赫阳
\end{CJK})$^{1}$, \
    Wanzhou Zhang(\begin{CJK}{UTF8}{gbsn}
张万舟
\end{CJK})$^{1,2}$ \thanks{Corresponding author. E-mail:zhangwanzhou@tyut.edu.cn} \
    ,Yanting Tian(\begin{CJK}{UTF8}{gbsn}
田彦婷
\end{CJK})$^{1}$,\ \\
    Chengxiang Ding(\begin{CJK}{UTF8}{gbsn}
丁成祥
\end{CJK})$^{3}$, \
    and Youjin Deng(\begin{CJK}{UTF8}{gbsn}
邓友金
\end{CJK})$^{2,4,5}$ \thanks{Corresponding author. E-mail:yjdeng@ustc.edu.cn}\\
    $^{1}${College of Physics, Taiyuan University of Technology, Shanxi 030024, China}\\  
    $^{2}${Hefei National Laboratory for Physical Sciences at the Microscale and Department of} \\{Modern Physics, University of Science and Technology of China, Hefei 230026, China}\\ 
    $^{3}${School of Science and Engineering of Mathematics and Physics,}\\
    {Anhui University of Technology, Maanshan, Anhui 243002, China}\\ 
    $^{4}${MinJiang Collaborative Center for Theoretical Physics, College of Physics and Electronic}\\ {Information Engineering, Minjiang University, Fuzhou 350108, China}\\ 
}   

\date{\today}
\maketitle

\begin{abstract}
    The two-component cold atom systems with anisotropic hopping amplitudes can be phenomenologically described by a two-dimensional Ising-XY coupled model with spatial anisotropy.
    At low temperatures, theoretical predictions [Phys. Rev. A 72, 053604 (2005)] and [arXiv:0706.1609] indicate the existence of a topological ordered phase characterized by Ising and XY disorder but with 2XY ordering. However, due to ergodic difficulties faced by Monte Carlo methods at low temperatures, this topological phase has not been numerically explored. We propose a linear cluster updating Monte Carlo method, which flips spins without rejection in the anisotropy limit but does not change the energy. Using this scheme and conventional Monte Carlo methods, we succeed in revealing the nature of topological phases with half-vortices and domain walls. In the constructed global phase diagram, Ising and XY type transitions are very close to each other and differ significantly from the schematic phase diagram reported earlier. We also propose and explore a wide range of quantities, including magnetism, superfluidity, specific heat, susceptibility, and even percolation susceptibility, and  obtain consistent and reliable results.
    Furthermore, we observe first-order transitions characterized by common intersection points in magnetizations for different system sizes, as opposed to the conventional phase transition where Binder cumulants of various sizes share common intersections. The results are useful to help cold atom experiments explore the half-vortex topological phase.
\end{abstract}



\section{Introduction}
In the 1970s, the topological phase and phase transitions were proposed and described by the two-dimensional XY model, which has broad applications including the description of superconductivity in 2D films, 2D crystals, 2D magnets, and various other systems~\cite{bkt_b,bkt_kt,bkt3,bkt4}. This phenomenon eventually is called  the celebrated Berezinskii-Kosterlitz-Thouless (BKT) transition, as referenced in the papers~\cite{bkt_b,bkt_kt}. The BKT transition typically involves the unbinding of vortex-antivortex pairs on a lattice~\cite{bktreview}. An integer vortex is defined as a circulation of the gradient of the angle $\theta$ of the standard XY spins, given by $\oint_{{ }}\nabla \theta \cdot d \vec \ell = 2\pi n$, where $n$ is an integer.

In addition to integer vortices, there are many non-integer vortices in topological phases, and the Ising transition emerges in such systems. Usually, half-integer vortex emerges in certain generalized XY models that include terms like $\text{cos}(2\theta_i- 2\theta_j)$~\cite{vortex_half}. These half-integer vortices are accompanied by strings or domain walls connecting pairs of them.
The Ising phase transition was observed due to the vanish of domain walls~\cite{vortex_half}. The so-called fully-frustrated XY model~\cite{ffxy1,isxy_1986}, and the equivalent version, i.e., Ising-XY coupled model~\cite{isxy1st,isxy2st,Granato1988HelicityMB} also has interaction between the domain wall and the vortex. Furthermore, the half-vortex and domain wall also exist in cold-atom systems~\cite{Multiflavor_bosonic_Hubbard}.

Due to the very close proximity of Ising and XY phase transitions and the presence of only one specific heat peak in some cases ~\cite{isxy1st,isxy2st}, distinguishing between the two phase transitions is challenging ~\cite{Olsson}. Numerous research efforts have been made to investigate the sequence in which these two types of phase transitions occur, with higher computational accuracy, using methods such as Monte Carlo simulations ~\cite{spin_chirality} or tensor network methods ~\cite{guangming}.

The Ising and XY transitions also occur in the  coupled Ising-XY model with spatially anisotropy~\cite{Multiflavor_bosonic_Hubbard, cenkexu}.
This model can be effectively derived from the two band Bose-Hubbard model that describes quantum gases. The XY spins are obtained through the mapping $a^{\dagger}\approx e^{-i\theta}$, where $a^{\dagger} (a)$ are the creation (annihilation) operators~\cite{Multiflavor_bosonic_Hubbard,cenkexu}.
This model was predicted to have an interesting topological phase denoted as the symbol $B_2$. The phase is characterized by Ising and XY disorder, but 2XY $(2\theta_1$ $+2\theta_2)$ order, where $\theta_{1(2)}$ are the XY spins of the layers, respectively.

However, it is difficult to simulate the model especially at low temperatures. The difficulty is the non-ergodic problems in Monte Carlo (MC) methods\cite{landau_binder_2014}. The $B_2$ phase is predicted Ising disordered~\cite{cenkexu}, but such disordered state cannot be reached if the initial state starts from an ordered state with all spins pointing in the same direction. At lower temperature, when using the Metropolis scheme~\cite{Metropolis1953EquationOS}, the states  become frozen in the energy local-minimum state and it is difficult to move the spins across the excited state to global ground state. Meanwhile, at low temperature, the Wolff clusters~\cite{wollf} becomes very large, and fail to  capture the true spin correlations, making cluster-updating inefficient.

In this paper, we have introduced innovations in {\it both} the method and the physics of the problem. Regarding the method, we propose a Monte Carlo (MC) scheme with line-shaped clusters, which can be either vertical or horizontal. In the case of spatial anisotropy, one can flip spins within the clusters without rejection while  keeping the energy invariant. This approach ensures consistent results  using different initial states. The reliability of our results is confirmed using the parallel tempering (replica exchange MC) method~\cite{pt}. Although the method looks simple, our scheme serves as a valuable reference for the study of other anisotropic models~\cite{gxy_potts, Multiflavor_bosonic_Hubbard}.

Regarding the physical aspects, while the model Hamiltonian and the topological phase ($B_2$) were initially proposed in Refs.~\cite{cenkexu, Multiflavor_bosonic_Hubbard}, our study has yielded distinct findings. These findings are summarized as follows:

\noindent (I) Our useful simulations reveal the presence of the $B_{2}$ phase by visualization of the strings and half vortices.
In addition to the fact that $\chi = \theta_1 + \theta_2$ is considered disordered~\cite{cenkexu}, we propose that $\theta_1$ itself should also be disordered.
The magnetization $M_{\theta_1}(L)$ follows a different way with the size in the topological phase (Ising disorder) and the high temperature disorder phase.
Another advantage is that the location of peaks in the distribution $P(M_{\theta_1})$ can be analyzed manually. This quantity and the quasi-one-dimensional pattern correspond very well in the anisotropic limit.

\noindent (II) In contrast to the schematic phase diagram, the numerical phase diagram shows two phase transition lines so close together that they are almost indistinguishable with the naked eye.
The distance between the two phase transitions is essentially on or less than the order of $10^{-2}$. For example, in the isotropic limit,
$T_c^{Ising}\approx 1.361(1)$ and
$T_c^{xy}\approx 1.342(4)$ are obtained.
In the anisotropic limit, the two phase transition boundaries are also very close to each other.
These values are in line with the findings reported in previous Refs.~\cite{Olsson,spin_chirality, guangming}.

\noindent (III) We have uncovered a first-order phase transition, which exhibits behavior distinct from that of systems with discrete spin variables, as described in the references~\cite{kbinder}. In typical cases, a phase transition is identified by the intersection points of the well-known Binder ratio for different system sizes~\cite{binder}. However, in our study, it is noteworthy that the magnetizations of different system sizes themselves share common intersection points.
This study provides a comprehensive understanding of the Ising-XY model with spatial anisotropy through reliable numerical investigations.

The outline of this paper is as follows. In Sec~\ref{sec:ham_phase}~~~~, the model and the predicted and simulated global phase diagram are shown. In Sec~\ref{sec:method_quantity}~~~~, the methods and various quantities are described.
The results are shown in Sec~\ref{sec:res}.
The nature of the $B_{2}$ phase and the phase transitions with other phases are discussed in detail.
Conclusion and discussion are made in Sec~\ref{sec:con}~~~~.

\section{Vortex, Model and global phase diagram}
\label{sec:ham_phase}
This section introduces the definitions of vortices and half-integer vortices, then describes the sources of Hamiltonian quantities for the specific spatially anisotropic coupled Ising-XY model, and finally gives a schematic representation of the phase diagram as well as numerical phase diagram obtained from our simulations.

\subsection{Vortex and anti-vortex}
\begin{figure}[htp]
    \centering
    \includegraphics[scale=0.5]{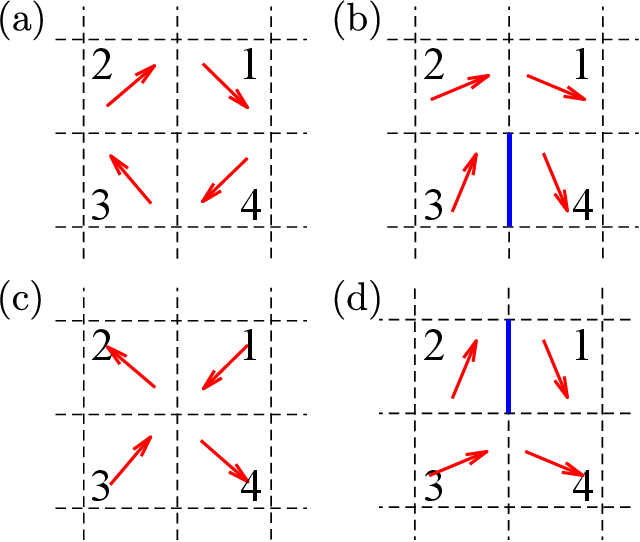}
    \caption{(a)    integer vortex, and (b)
        half-integer vortex
        and (c)    integer anti-vortex, (d) half-integer anti-vortex.
        The blue line is the domain wall between two spins in a half-integer vortex with an angular difference close to $\pi$.}
    \label{fig:vortex}
\end{figure}

The integer vortices, half-integer vortices and mathematical descriptions are given here before the specific physical models are presented. In Fig.~\ref{fig:vortex} (a), the integer vortex  is shown. Mathematically, it is defined by the four spins around a plaquette $\square{ }$.
The four spin angles $\theta_i$, $i=1,2,3,4$,  can be restricted to [$-\pi$, $\pi$] by  $\pm 2\pi$.
The sum of the differences between adjacent spins  is defined as
$2\pi n = (\theta_2 - \theta_{1})  + (\theta_3 - \theta_{2})
    + (\theta_4 - \theta_{3}) + (\theta_1 - \theta_{4})$. $2\pi n $ is naturally equal to 0 if four values in the brackets $( )$ are in the range (-$\pi$, $\pi$].
However, whenever any of these values in brackets  exceeds the range, there is a possibility of nonzero vorticity by the saw-tooth function, defined by~\cite{roger}
\begin{equation}
    \label{eq:saw}
    \text{saw}( \theta_i- \theta_{j} ) = \begin{cases}
        \theta_i- \theta_{j} + 2 \pi, & \theta_i- \theta_{j}\leq -\pi,        \\
        \theta_i- \theta_{j},         & -\pi < \theta_i- \theta_{j} \leq \pi, \\
        \theta_i- \theta_{j} - 2\pi,  & \pi < \theta_i- \theta_{j}.
    \end{cases}
\end{equation}
When using the configurations shown in Fig.~\ref{fig:vortex} (a) with $\theta_{1,2,3,4}=-\pi/4$, $\pi/4$, $3\pi/4$, $-3\pi/4$, one can get $\Delta \theta_{i,j}=\pi/2$, $\pi/2$, $-3\pi/2$, $\pi/2$, respectively. Without using the saw-tooth function, the summation becomes $\sum \Delta \theta_{i,j}=0$.
However, when the saw-tooth function is employed, the summation results in $\sum \Delta \theta_{i,j}=2\pi n, n=1$.

In Fig.~\ref{fig:vortex} (b), the angles are set to be $\theta_{1,2,3,4}=-\pi/8$, $\pi/8$, $3\pi/8$, $-3\pi/8$, respectively. This configuration results in angle differences of  $\Delta \theta_{i,j}=\pi/4$, $\pi/4$, $-3\pi/4$, $\pi/4$. Using the saw-tooth function with  the modified bounding range  $(-\pi/2$ , $\pi/2]$,  the summation   yields $\sum \Delta \theta_{i,j}=2\pi n$, where  $n=1/2$.
In other words, the spin  plaquette corresponding to the half-vortex in the configuration $\{\theta_i\}$ is also the integer vortex  plaquette corresponding to the $\{2\theta_i\}$ configuration~\cite{diss}.

{Note that the saw-tooth function is used to detect the presence of half-vortices rather than causing them, indicating its independence from the physics of the Ising-XY model.}

\subsection{Models}

The two component Bose-Hubbard model \cite{cold_Bosonic_atom,cenkexu,Multiflavor_bosonic_Hubbard}, which describes cold atoms with $p_x$ and $p_y$ orbitals, can be mapped to the double-layer XY model~\cite{cenkexu}. The Hamiltonian of this model is composed of three terms: $H = H_{up} + H_{dn} + H_{updn}$, where $H_{up}$, $H_{dn}$, and $H_{updn}$ represent the interaction of spins within the upper layer, within the lower layer, and between the two layers, respectively. These terms are given by :
\begin{align}
    H_{up}   & = \sum_{i,j} \left[-J_a \cos \left(\nabla_x \theta_{i,j, 1}\right) - J_b \cos \left(\nabla_y \theta_{i,j, 1}\right) \right],\notag   \\
    H_{dn}   & =\sum_{i,j}\left[-J_b \cos \left(\nabla_x \theta_{i,j, 2}\right) - J_a \cos \left(\nabla_y \theta_{i,j, 2}\right)\right],     \notag \\
    H_{updn} & =\sum_{i,j}\left[\gamma \cos \left(2 \theta_{i, j,1} - 2 \theta_{i,j, 2}\right)\right] .
    \label{eq:ham}
\end{align}
Here, $J_a$, $J_b$, and $\gamma$ represent the interaction strengths. The indices, denoted as $i$ and $j$, vary from 1 to $L$ during the summing, where $L$ represents the size of the square lattice. Additionally, the total number of XY spins in the upper (lower) layer is $N = L\times L$. $\theta_{i,j,1(2)}$ represents the XY spin variables on the lattice site $(i,j)$ at the upper (lower) layer.
It is proposed that the XY spins between the two layers are perpendicular to each other.  
Eqs.~\ref{eq:ham} are not artificially created Hamiltonians as the Bose Hubbard model can describe real quantum gases, 

To account for this perpendicular alignment, the Ising variable $\sigma$ is introduced, satisfying the relation
\begin{equation}
    \theta_2 = \theta_1 + \sigma \pi/2.
    \label{eq:map}
\end{equation}
One substitutes this relation into the Hamiltonian, disregarding the constant term $H_{updn}=-\gamma N$, where $N$ also represents the number of spin pairs between the two layers.
The model thus obtained is the Ising-XY model with spatial anisotropy, given by the following equation:
\begin{align}
    H & = \sum_{ i,j } -\left(J_b + J_a \sigma_{i,j} \sigma_{i+\vec{x},j}\right) \cos(\theta_{i,j} - \theta_{i+\vec{x},j}) \nonumber\ \\
      & \quad + \sum_{ i,j } -\left(J_a + J_b \sigma_{i,j} \sigma_{i,j+\vec{y}}\right) \cos(\theta_{i,j} - \theta_{i,j+\vec{y}}),
    \label{eq:ixy}
\end{align}
{where $\theta\in[-\pi,\pi]$, and  the angular difference $\Delta\theta=\theta_{i,j} - \theta_{i+\vec{x},j}$ does not rely on the saw-tooth function.
    This function is specifically employed for detecting vortices but does not modify any expression in the Hamiltonian.
}
$\sigma_{i,j}$ represents the Ising spin variables on the lattice site $(i,j)$, and $\sigma_{i,j} \sigma_{i+\vec{x},j}$ represents interactions along the $x$ direction.
In particular, $\theta_1$ denotes the spin angle in the bilayer XY model before the mapping, and $\theta$ denotes the spin angle in the Ising-XY model after the mapping Eq.~\ref{eq:map}.

\begin{figure}[tb]
    \centering
    \includegraphics[scale=0.46]{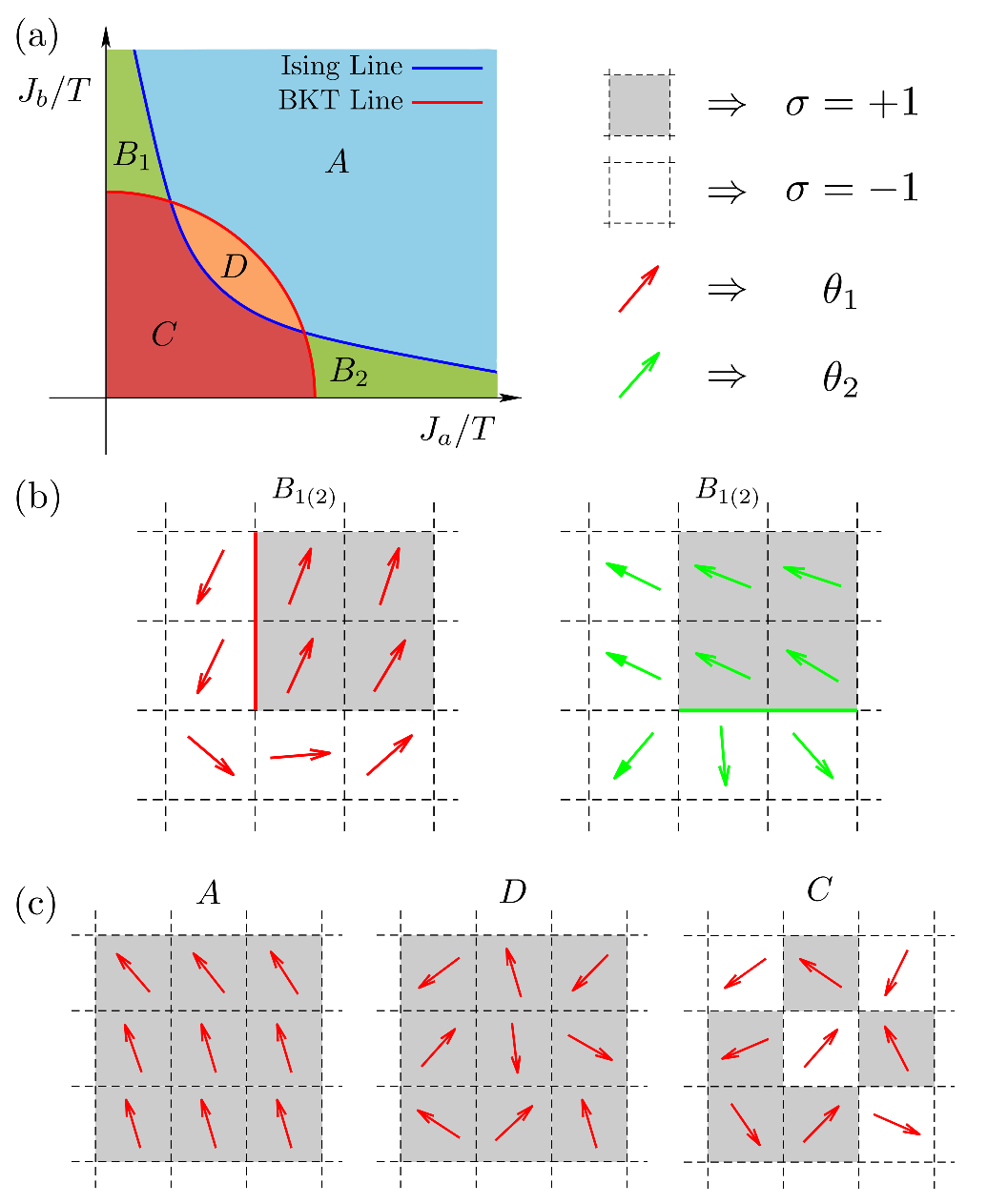}
    \caption{(a) Schematic phase diagram  of the model in
        Eq.~\ref{eq:ixy} and (b) typical configurations $\{\theta_1\}$ and $\{\theta_2\}$ of the half vortex in the phases $B_{1}$ or $B_{2}$ presented  by Ref.~\cite{cenkexu}.
        (c) Typical configurations of the phase $A$, $D$ and $C$. Red arrows mean the XY spins $\theta_1$, gray and white squares represent the Ising spins ($\sigma=1$ and $-1$). The XY spins $\theta_2$ can be obtained by the mapping in Eq.~\ref{eq:map}.}
    \label{fig:xucenke}
\end{figure}

\subsection{Phase diagram}
Figure~\ref{fig:xucenke}(a) shows the schematic phase diagram of the model in Eq.~\ref{eq:ixy}~\cite{cenkexu}. It contains the  four phases $A$, $B_{1(2)}$, $C$, and $D$. The phase diagram is symmetric about the $J_a=J_b$ axis in the plane $J_a/T-J_b/T$. In order to demonstrate the configurations of each phase,  the gray square represents the upward-oriented Ising spin ($\sigma_i = 1$), while downwards spins are represented by the color white ($\sigma_i = -1$). The red arrows indicate the XY spins $\theta_1$ in the upper layer.  The green arrows indicate the XY spins $\theta_2$ in the lower layer.

Figure~\ref{fig:xucenke}(b) illustrates two typical configurations in the $B_2$ phase as illustrated in Refs.~\cite{half_bh,cenkexu}.  The vertical dashed red line  denotes the domain wall of the XY spin $\theta_1$, while the XY spin in the other layer, $\theta_2$ exhibits domain walls in the horizontal direction through  Eq.~\ref{eq:map}.
The main characteristic of the $B_2$ phase is the absence of Ising order (sum of Ising variables) due to the presence of numerous domain walls in the actual simulations. Similarly, the long-range XY spin order, as seen in the configurations of $\chi=\theta_1+\theta_2$, is also expected to be disorder due to the domain walls for the XY spins. However, it is predicted that the $B_2$ phase exhibits ordered for the configuration of $2\chi=2\theta_1+2\theta_2$.
To provide a clearer understanding of the above terminology, let's consider the following example.  Suppose there are two spins with opposite orientations on each side of the domain wall, i.e., $\theta_{i,j}-\theta_{i+1,j}\approx\pi$,
according to Ref.~\cite{cenkexu}, doubling each angle magnifies the angular difference to $2\pi$, i.e., $2\theta_{i,j}-2\theta_{i+1,j}\approx2\pi$. Consequently, for spins pointing in opposite directions, doubling their angles results in them becoming identically oriented.
The above analysis also applies to the domain wall in  the $x$-direction for $\theta_2$ (not shown).
In summary, the phase $B_2$ is predicted to behave in both Ising and XY ($\chi=\theta_{1}+\theta_{2}$) disorder but 2XY ($2\chi$) ordered.

Figure~\ref{fig:xucenke} (c) depicts configurations of the phases $A$, $D$, and $C$, respectively. The phase $A$ corresponds to an ordered configuration of both Ising and XY spins, the phase $D$ exhibits the Ising order but XY disorder, and the phase $C$ represents disorder in both Ising and XY spins. The  order parameters for these four phases are listed in Table.~\ref{tab:my_label}.

\begin{table}[htp]
    \centering
    \caption{Different phases and their order parameters}
    \label{tab:my_label}
    \begin{tabular}{lccccc}
        \hline
        phases/orders       & ~~$A$~~    & ~~$C$~~      & ~~$D$~~      & ~~$B_{1}$~~  & ~~$B_2$~~    \\
        \midrule
        Ising order         & \checkmark & \XSolidBrush & \checkmark   & \XSolidBrush & \XSolidBrush \\
        XY order ($\chi$)   & \checkmark & \XSolidBrush & \XSolidBrush & \XSolidBrush & \XSolidBrush \\
        2XY order ($2\chi$) & \checkmark & \XSolidBrush & \XSolidBrush & \checkmark   & \checkmark   \\
        \bottomrule
    \end{tabular}

\end{table}

Moreover, in the thermodynamic limit, $L\rightarrow\infty$,  we propose that $M_{\theta}$ also tends to 0 in the $B_2$ phase.
However,  the behavior of  $M_{\theta}(L)$ as a function of  $L$ exhibits distinct trends in the $B_2$ phase and the high-temperature disorder $C$ phase.
This is discussed  in more detail in  Sec~\ref{sec:jb<<ja}~~~~.

\begin{figure}[t]
    \centering
    \includegraphics[scale=0.6]{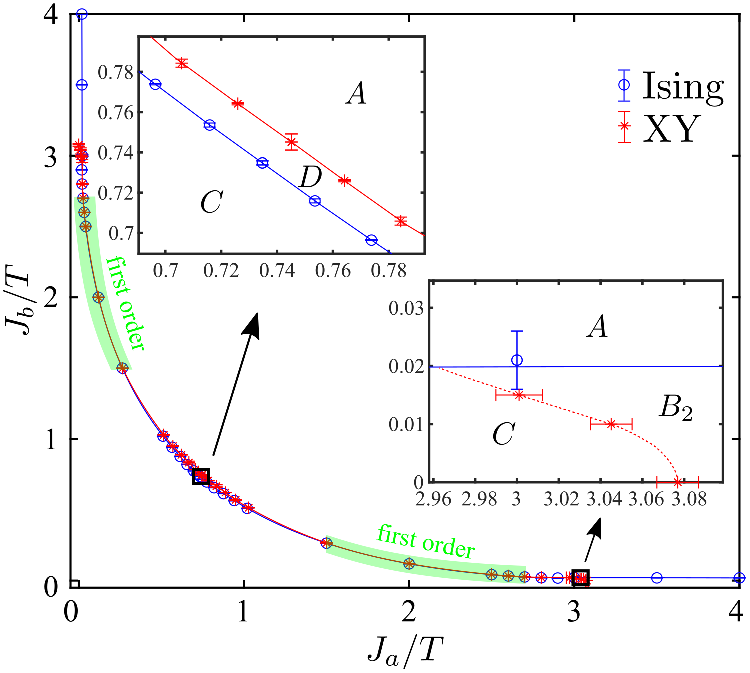}
    \caption{Global phase diagram obtained by the Monte Carlo method.   First order phase transitions are found $J_{a}/T = 1.5-2.7$.
        The insets are to make the phase transition boundary visible to the naked eye.
    }
    \label{fig:global}
\end{figure}
In Fig.~\ref{fig:global}, we obtain the global phase diagram by the numerical MC method.
In the  plane $J_a/T-J_b/T$, as shown in Fig.~\ref{fig:xucenke}, the topology of the phase diagram is consistent with the results predicted by the theory. The phase diagrams still have the $A$, $B_1(B_2)$, $C$, and $D$ phases.

However, as presented  by the numerical simulation results, the Ising and XY phase transitions are very close to each other.
The change in the sequence of the Ising and XY phase transitions is confirmed. It is difficult to distinguish the two phase transition lines with the naked eye, and for this reason we have drawn enlarged pictures in certain areas, which are placed in the two insets. One inset shows the positions of the $A$, $D$ and $C$ phases.
The other inset shows the positions of the $A$, $C$ and $B_2$ phases.
In the regime $J_b/T=0.1, J_a/T=1.5-2.7$ marked by green  in Fig.~\ref{fig:global},
first order transitions are found.  Other details such as the snapshots of the $B_{1(2)}$ phases are discussed in Sec~\ref{sec:res}~~~~.

\section{Methods and Observed quantities}
\label{sec:method_quantity}

In order to  construct  the global phase diagram,
we simulate the model in the various values of $J_b/J_a$.
\sethlcolor{green}{
    We primarily employ the Metropolis-Wolff algorithm in the regime of $J_b/J_a \in [0.2, 1]$, while for the regime where $J_b/J_a<0.2$, we utilize the Metropolis-Wolff-Line algorithm. The results obtained through this approach have been cross-validated using the parallel tempering (PT) MC method~\cite{pt}.}
\subsection{\textcolor{black}{Line-shaped cluster-updating Method}}
\label{sec:method}


\begin{figure}[t]
    \centering    \includegraphics[scale=0.5]{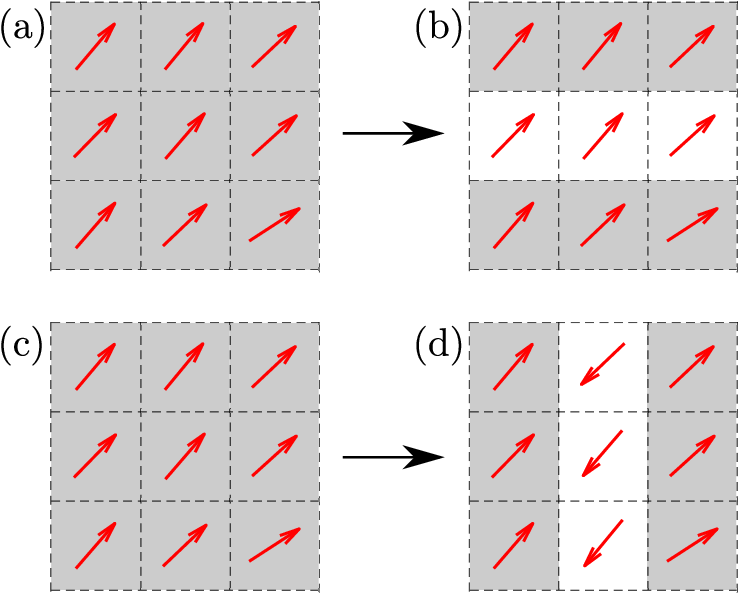}
    \caption{Schematic diagram of line-shaped cluster update.
        (a) and (b) update along the $x$-direction with flipping only the Ising spins; (c) and (d) update along the $y$-direction, flipping both the Ising spins and rotating the angle of the XY spin $180^{\circ}$.}
    \label{line_cluster_update}
\end{figure}

\subsubsection{\textcolor{black}{The energy-constant updating with line-shaped cluster}}

The basic idea of the line-shaped cluster MC is as follows. In the parameter limit $J_{b}= 0$, one can flip randomly without rejecting the spins within the clusters as the energy does not change.
The energies of the configurations in Fig.~\ref{line_cluster_update} (a) and Fig.~\ref{line_cluster_update} (b) are the same by substituting $J_b = 0$ into Eq.~\ref{eq:ixy}.

Similarly, in the second line of  Eq.~\ref{eq:ixy}, in the $y$-direction, one  rotates  the XY spins with angles $\pi$ simultaneously. To fix the energy term  in the $x$-direction, one has to flip the Ising variables simultaneously. Specifically, this can be illustrated by the following equation
\begin{equation}
    \begin{split}
        J_{a}\sigma_{i,j}\sigma_{i+\vec{x},j}\cos(\theta_{i,j}-\theta_{i+\vec{x},j}) \equiv \\J_{a}(-\sigma_{i,j})\sigma_{i+\vec{x},j}\cos([\theta_{i,j}+\pi]-\theta_{i+\vec{x},j}).
    \end{split}
\end{equation}
The advantage of this scheme is that there is no probability of rejection, and it can help us to explore the $B_2$ phase at $J_b = 0$ very well.

\textcolor{black}{\subsubsection{The updating of a line-shaped cluster with changes in energy}}

In the parameter regime $J_{b}\ne 0$ but very close to $0$. One can use the Metropolis method to accept or reject the flip of the spins in the above line-shaped clusters defined as follows:
\begin{align}
    \Delta E_{1} & =\sum_{i=1}^{L}\sum_{y=-1}^{\pm1}2J_{b}\sigma_{i,j}\sigma_{i,j+y}\cos{(\theta_{i,j}-\theta_{i,j+y})}\nonumber \\
    \Delta E_{2} & =\sum_{j=1}^{L}\sum_{x=-1}^{\pm1}2J_{b}\cos{(\theta_{i,j}-\theta_{i+x,j})}\nonumber                           \\
    p_{accept}   & = \left\{\begin{matrix}
                                exp(-\beta\Delta E_{1}) & \quad(\text{Fig.~\ref{line_cluster_update}(a) $\rightarrow$ (b)}) \\
                                exp(-\beta\Delta E_{2}) & \quad(\text{Fig.~\ref{line_cluster_update}(c) $\rightarrow$ (d)})
                            \end{matrix}\right.
\end{align}

In the above equation,  $p_{accept}$ is acceptance probability, $\Delta E_{1}$, $\Delta E_{2}$ are the energy difference by fliping the clusters along $x$ and $y$ direction respectively.
In the real simulation, we perform $L$ times in each MC step to select the directions of the line clusters, which are oriented either horizontally or vertically with a probability of 0.5 and then try to flip the spins within the clusters.
\subsubsection{\textcolor{black}{One MC step with three kinds of updating}}

{
    In the real simulation, in one MC step, we use  three kinds of updating at the same time.
    These updates include the line-shaped cluster updating, the Metropolis algorithm~\cite{Metropolis1953EquationOS}, and   the Wolff clustering algorithm~\cite{wollf}.
    The constant-energy flips, without rejections, sample a subset of phase space confined to a constant-energy surface. Therefore, we let the configurations jump out of the subset of energy constants with the help of the Metropolis and Wolff algorithms. The idea of mixing up different algorithms is used in other references~\cite{landau_binder_2014,PhysRevB.108.024402,PhysRevB.105.224415}.
}
\subsection{Parallel Tempering method}
Parallel Tempering (PT) is a simulation technique that enhances the efficiency of MC sampling methods~\cite{pt}, especially useful in spin-glass systems~\cite{rmp_glass,pt_glass2,PT3}. It is particularly effective for systems with complex energy landscapes, where traditional sampling methods may get trapped in local minima.
{The basic idea behind the PT method is to simulate multiple replicas  at different temperatures simultaneously. Each replica represents a copy of the system with the same set of spins but at a different temperature.} The acceptance probability of an exchange between neighboring replicas satisfies
\begin{figure}[tb]
    \centering
    \includegraphics[scale=0.75]{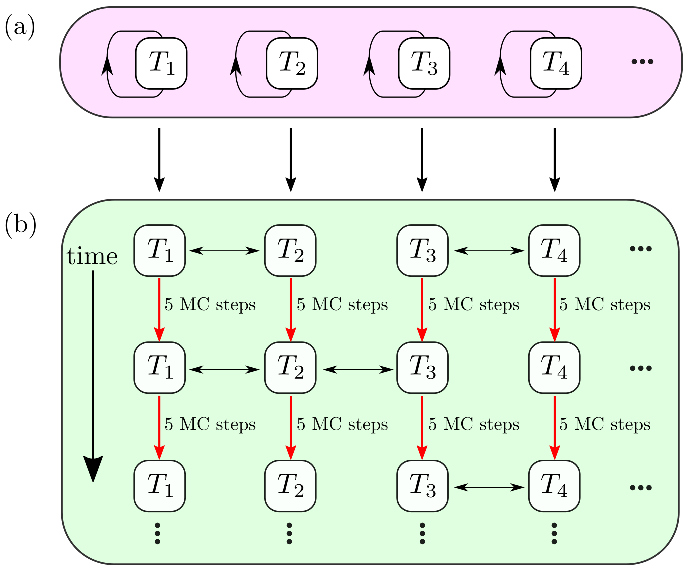}
    \caption{\textcolor{black}{Parallel tempering involves multiple processes, each denoted as $T_i$ where $i = 1, 2, \dots$, representing replicas at different temperatures. (a) the independent relaxation for each replica, (b)  the process of exchanging replicas. Bidirectional arrows indicate successful replica exchanges, and red arrows indicate that 5 MC steps are simulated between exchanges.}}
    \label{fig:PT}
\end{figure}
\begin{equation}
    \begin{split}
        p       &=\min \left(1, e^{\left(E_{i}-E_{j}\right)\left(\frac{1}{ T_{i}}-\frac{1}{T_{j}}\right)}\right)
    \end{split},
    \label{eq:pb}
\end{equation}
{where $T_{i(j)}$ represents the temperature of the replica, and $E_{i(j)}$ is the energy  at temperature $T_{i(j)}$.} In the actual simulation, we set the temperature range to cover both sides of the phase transition. This approach allows the low-temperature system to feel the influence of the high-temperature system and ensures that the system explores the entire energy landscape, including the local energy minimum.

{The  two stages of the PT method are described in detail. In Fig.}~\ref{fig:PT} { (a),  taking $J_b=0$ for example, for the relaxation stage,  250 discrete temperatures are uniformly taken in the range [0,1]. In each $T_i$, 500,000 relaxation MC steps are performed, and one MC step consists of Wolff-Metropolis-Line updates. In Fig.}~\ref{fig:PT} (b), {during the exchange stage, 2,500,000 MC updates are performed for each $T_i$. Every 5 MC steps, configurations at each $T_i$ are attempted to be exchanged with configurations at neighboring temperatures $T_{i-1}$ and $T_{i+1}$ with the acceptance probability given }in Eq.~\ref{eq:pb}.
The bidirectional arrows indicate successful replica exchanges.

\subsection{Algorithmic verification}
\begin{figure}[tb]
    \centering
    \includegraphics[scale=1]{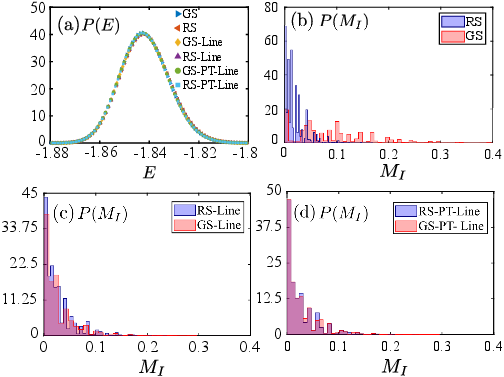}
    \caption{Distribution of energy  $P(E)$ and magnetization  $P(M_{I})$ for different methods and different initial configurations.
        (a) The same $P(E)$ by  three methods with two different  initial configurations.
        (b) $P(M_{I})$ by the Wolff-Metropolis method, (c) $P(M_{I})$ by the Wolff-Metropolis-Line method, and (d) $P(M_{I})$ by the Wolff-Metropolis-Line-PT method.
    }
    \label{fig:compare}
\end{figure}

In Fig.~\ref{fig:compare}, we compare the effects of the combination of different methods at low temperatures and in the anisotropic limit.
There are three methods in total, one of which is a pure Wolff-Metropolis method. In certain cases, if the Wolff cluster is large, the MC algorithm may fail to meet the ergodicity condition. In such situations, several Metropolis-type updates can be employed to address this issue.
The second method is the Wolff-Metropolis-Line method, and the third is the Wolff-Metropolis-Line-PT method.
The three methods utilize different initial configurations, random  or all the same, i.e., marked by "RS" and "GS".
With parameters $L=24$, $J_{b}=0$, $T=0.3$, for the distribution of the energy and Ising magnetization are measured.

In Fig.~\ref{fig:compare} (a), the distribution of energy $P(E)$ is the same for all methods with different initial states.
The energy probability distributions are identical in all cases, but the magnetization probability distributions are different in the Fig.~\ref{fig:compare} (b), indicating that sampling does not traverse the global configuration space well, and only samples certain degenerated states of the energy state.

As seen in Fig.~\ref{fig:compare} (c), the Wolff-Metropolis-Line method allows a high degree of overlap between $P(M_{I})$ obtained from the two different initial states.
The minor deviation observed between the two distributions is primarily caused by fluctuations. Finally, in  Fig.~\ref{fig:compare} (d), the Wolff-Metropolis-Line-PT method confirms the convergence of the distribution $P(M_{I})$, so there is reason to believe that better global sampling is achieved.

\subsection{Observed quantities}

(\Rmnum{1}) Both the BKT and  Ising phase transitions can be described by the magnetic order parameters, defined by the Ising  and the XY magnetization defined as,
\begin{align}
     & M_{\text{I}}=\frac{1}{N}\expval{\abs{\sum_{i}\sigma_{i}}},                                                                                 \\
     & M_{\theta}=\frac{1}{N}\left\langle\sqrt{\left(\sum_{j} \sin \theta_{j}\right)^{2}+\left(\sum_{j} \cos \theta_{j}\right)^{2}}\right\rangle,
\end{align}
where $\sigma=\pm 1$ represents the Ising spin,  and $N$ represents the total number of lattice sites.
$M_{\theta}$ is used to define the BKT phase transition with respect to $\theta$, similarly, $M_{\chi}$ and $M_{2\chi}$ can be defined as magnetization order parameters by the variables $\chi$ and $2\chi$. In the calculations of $M_{\chi}$ and $M_{2\chi}$, neither $\chi$ nor $2\chi$ is subject to regulation by the saw-tooth function. This is because the quantities $M_{\chi}$ and $M_{2\chi}$ are employed for detecting magnetism, not vortices.
Note that $\theta_2=\theta_1+\sigma\frac{\pi}{2}$, and $\chi=\theta_1+\theta_2$.

(\Rmnum{2}) The Binder  cumulant~\cite{binder,sandvik} for the Ising and XY variables are defined as,
\begin{equation}
    U_{2}=\frac{n+2}{2}\left(1-\frac{n}{n+2}R_{2}\right),
    \label{eq:binder}
\end{equation}
where $n=1$ for the Ising variables, $n=2$ for the XY variables, and $R_2$ is Binder ratio~\cite{PhysRevB.30.1477, PhysRevLett.47.693} defined as
\begin{equation}
    R_{2}=\expval{M^{4}}/\expval{M^2}^2.
\end{equation}
(\Rmnum{3}) Specific heat and susceptibilities. We introduce corresponding susceptibilities for all order parameters ($M_{I}, M_{\theta}, M_{\chi}, M_{2\chi}$),
\begin{equation}
    \chi_s=\frac{1}{k_B T}(\expval{M^2}-\expval{M}^2),
    \label{eq:sus}
\end{equation}
and the specific heat $C_V$,
\begin{equation}
    C_V=\frac{1}{k_B T^2}(\expval{H^2}-\expval{H}^2).
\end{equation}

(\Rmnum{4}) The spin stiffness  $\rho_{s}$ is defined as follows~\cite{sandvik}:
\begin{eqnarray}
    \mathcal{\rho}_{s}=\frac {1} {Nd}[\langle H \rangle-\beta(\langle I^2_{x} \rangle + \langle I^2_{y} \rangle)],
    \label{eq:sf}
\end{eqnarray}
where $\beta$ is the inverse temperature, $N$ is the total number of spins, $d$ is the dimension, and $I_{x}, I_{y}$ defined as
\begin{align}
    I_{x}= & \sum_{<i,j>_{x}}(J_{b}+J_{a}\sigma_{i,j}\sigma_{i+\vec{x},j})\sin(\theta_{i,j}-\theta_{i+\vec{x},j}), \\
    I_{y}= & \sum_{<i,j>_{y}}(J_{a}+J_{b}\sigma_{i,j}\sigma_{i,j+\vec{y}})\sin(\theta_{i,j}-\theta_{i,j+\vec{y}}).
    \label{eq:sfi}
\end{align}

(\Rmnum{5}) The bond's order in $x$- and $y$-directions
\begin{align}
     & Bond_{x}=\sum_{<i,j>_{x}}\sigma_i\sigma_j,
     & Bond_{y}=\sum_{<i,j>_{y}}\sigma_i\sigma_j,
    \label{eq:bond}
\end{align}
where are useful to characterize the melting Ising domain walls.\\

(\Rmnum{6}) The percolation susceptibility \cite{PhysRevE.79.061118} $S_{2}$ is defined as
\begin{equation}
    S_{2}=L^{-2}\expval{  \sum_{i=1}^{N_{c}-1}n_{i}^{2}  },
    \label{eq:perco}
\end{equation}
\sethlcolor{green}{where $n_i$ denotes the number of Ising spins in the Ising clusters}. Meanwhile, one removes the  cluster containing the highest number of sites during the summing.

\section{Results}
\label{sec:res}
In this section, we give numerical results in detail.
We first scan the phase diagram along the $J_a = J_b$ diagonal cut in Sec~\ref{sec:jajb}~~~~, then we scan the regimes {$J_b\ll J_a$} in Sec~\ref{sec:jb<<ja}~~~~.
The signature of $M_{\theta}$ is also analyzed carefully in Sec~\ref{sec:mtheta}~~~~.
The regimes of first order transition is confirmed in range about $J_a/T\in (1.5,2.7)$ in Sec~\ref{sec:1st}~~~~.
At $J_a/T=3.5$, the phase transition and Ising type universality  between $A-B_2$ phase are discussed in Sec~\ref{sec:Ja=3.5}~~~~.

\subsection{Phase transitions between the phases
\texorpdfstring{$A-D-C$}{}  along \texorpdfstring{$J_{b}=J_{a}$}{}}
\label{sec:jajb}
In this section, we calculate the critical points of the Ising transition and the XY  transition at isotropic parameters $J_{b}=J_{a}$ and verify that the XY  transition occurs earlier than the Ising  transition from low to high temperature. It is understandable that if the Ising spins emerge as a domain wall, the interaction $(1+\sigma_{i}\sigma_{j})=0$,
so the algebraic order of the XY spins can only exist before the Ising  transition or both the XY and Ising transitions occur at the same time.
In addition, in the coupled system, we also confirm the universalities  by data collapse method on the data with different lattices.
\begin{figure}[t]
    \centering
    \includegraphics[scale=0.4]{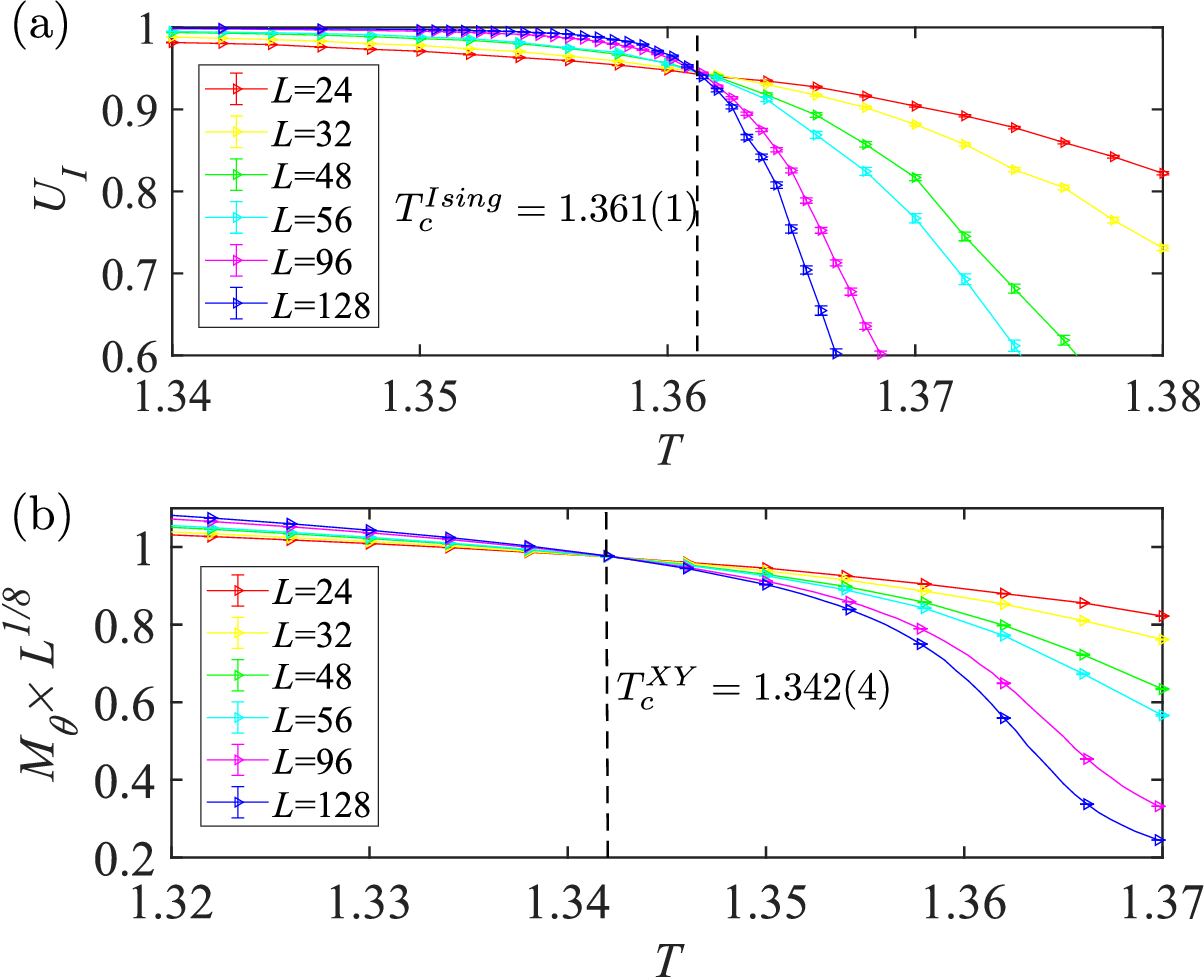}
    \vspace{3mm} \\
    \includegraphics[scale=0.38]{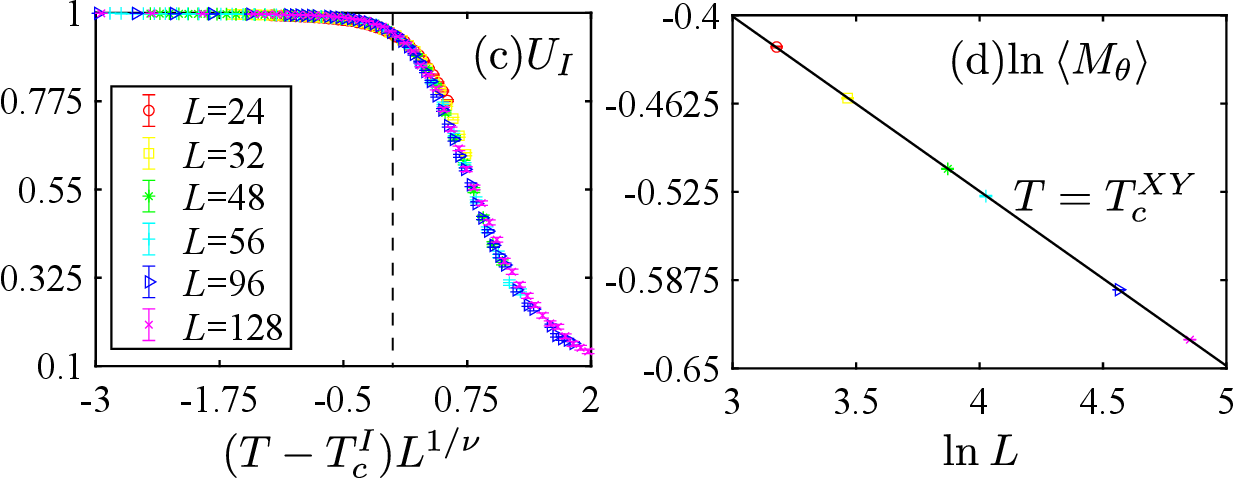}
    \caption{The results  along the cut $J_a=J_b$. (a) the crossing points of $U_I$  occur at $T_c^{Ising}\approx 1.361(1)$, (b) the crossing points of $M_{\theta}L^{\frac{1}{8}}$  occur at
    $T_c^{xy}\approx 1.342(4)$. (c) Data collapse of $U_I$ (d)  $\ln M_I$ .vs. $\ln L.$
    }
    \label{fig:A=B_phase_transition}
\end{figure}

In Fig.~\ref{fig:A=B_phase_transition} (a),  along the diagonal cut $J_a=J_b$ in the phase diagram, we scan the temperature in the range $T\in [1.34,1.38]$ with lattice sizes $L=24, 32, 48, 56, 96, 128$.
With the increase in temperature, the Ising binder cumulative $U_I$ overlap at $T_{c}^{I}\approx 1.361(1)$ for different lattices.
At critical regimes, the cumulative moments of the magnetization satisfy the following scaling relationship as~\cite{binder, sandvik_ling},
\begin{equation}
    \langle M^{k} \rangle_L = L^{-k\beta/\nu} \mathcal{F}(\frac{T-T_c}{T_c} L^{1/\nu}),
    \label{eq:mk}
\end{equation}
where $\mathcal{F}$ is a scaling function. When $T=T_c$, the ratio $R_2=\langle M^{4} \rangle_L/\langle M^{2} \rangle_L^2$ is independent of $L$, and therefore the data $U_I$ for different sizes cross at the critical point.

In Fig.~\ref{fig:A=B_phase_transition} (b), $M_{\theta}L^{1/8}$ vs $T$ is plotted. The data from different sizes also cross at $T=T_c$ very well. The reason is as follows. Firstly, the correlation length is proportional to  system size as
\begin{equation}
    \xi (T) \sim L \sim \lvert \frac{T-T_c}{T_c} \rvert ^{-\nu},
\end{equation}
and then
\begin{equation}
    M_{\theta
    } \sim  \lvert \frac{T-T_c}{T_c} \rvert ^{\beta} \sim L^{-\beta/\nu}.
\end{equation}
Using the critical exponents $\beta = 1/8$ and $\nu=1$, the data of $M_{\theta}L^{1/8}$ cross at the same point for different sizes.
$M_{\theta}L^{1/8}$ is a very good order parameter to locate the BKT phase transition point $T_{c}^{XY}\approx 1.342(4)$.

To further confirm the Ising transition, we also re-scale  $T$ as $(T-T_c)L^{1/\nu}$, and the data overlap very well as shown in Fig.~\ref{fig:A=B_phase_transition} (c).
For the XY transition, letting $k=1$ in Eq.~\ref{eq:mk}, one can get $\langle M_{\theta} \rangle_L \propto L^{-1/8}$. Using the log-log plot, the slope is -1/8 as shown in Fig.~\ref{fig:A=B_phase_transition} (d).

\subsection{Snapshots, and phase transition between the phases \texorpdfstring{$B_2-C$}{} with \texorpdfstring{$J_b\ll J_a$}{}}
\label{sec:jb<<ja}
\begin{figure*}[htp]
    \centering
    \includegraphics[scale=0.6]{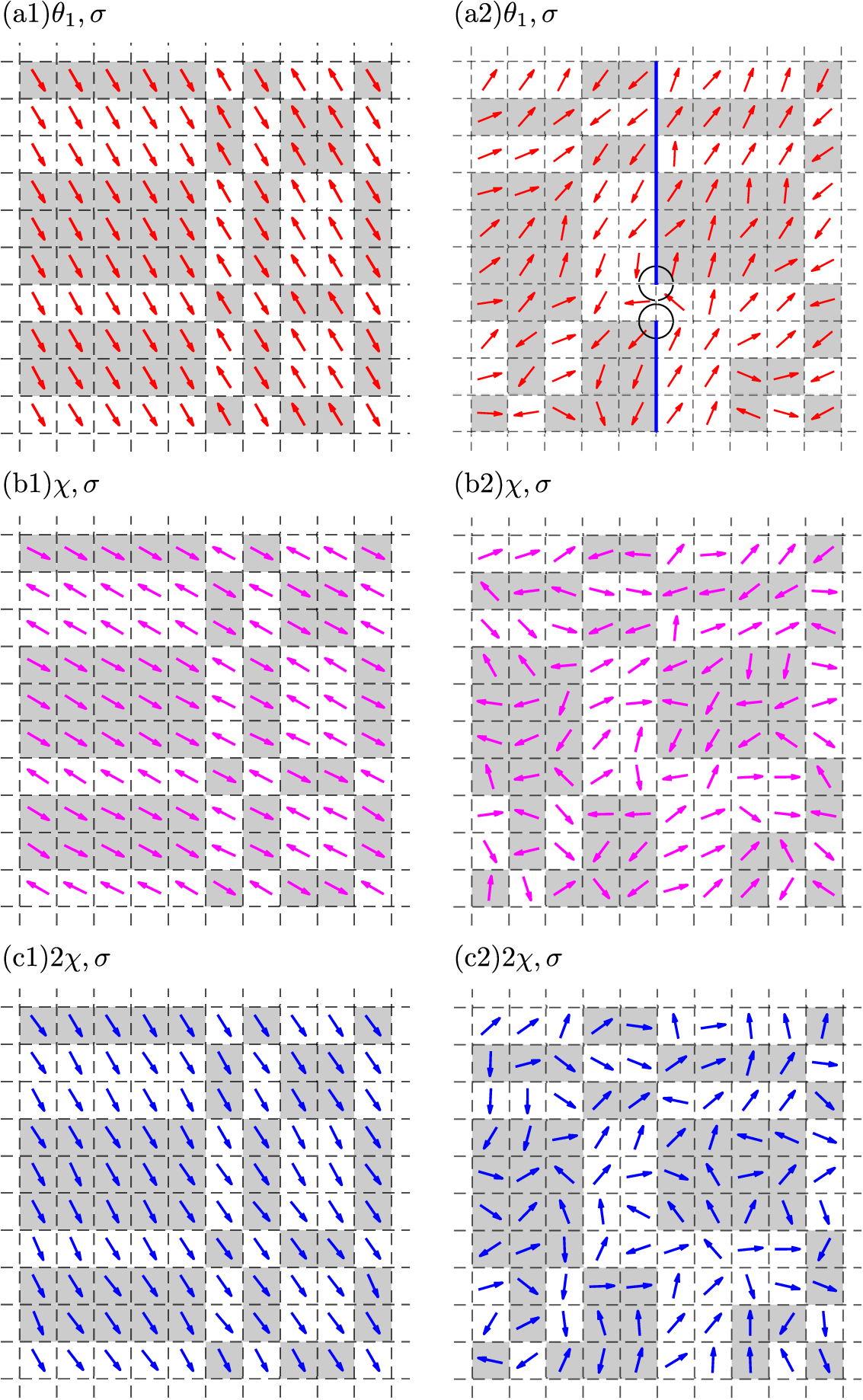}
    \caption{ Snapshots for $\{\theta_1$, $\sigma$, $\chi$, $2\chi\}$. Left column: $J_{a}=1$, $J_{b}=0$, $T=0.01$ in the $B_{2}$ phase. Right column: $J_{a}=1$, $J_{b}=0$, $T=0.375$ near the transition point.  The Arrows indicate the XY spins, and the shaded squares indicate the Ising spins $\sigma_i=1$, and the white squares indicate $\sigma_i=-1$.
        (a) The snapshots of $\{\theta_1\}$ and $\{\sigma\}$. The domain walls of $\theta_1$ (red)
        form straight lines aligned in the $y$-direction while the Ising domain walls are represented as closed rectangles.
        (b) The snapshots of ${\chi=\theta_1+\theta_2}$ clearly reveal the disorder in $\chi$ with the magnetization  $M_{\chi}\approx 0.079$ (left) and $M_{\chi}\approx 0.006$ (right).
        (c) The snapshots of  $\{2\chi\}$  demonstrate the ordered phase, with $M_{2\chi}\approx 0.998$ (left) and $M_{2\chi}\approx 0.367$ (right, near the critical point).
    }
    \label{fig:jb=0snapshot}
\end{figure*}
Figure~\ref{fig:jb=0snapshot} plots typical snapshots in the $B_2$ phase  ($J_b/J_{a}=0$), showing the  configurations of $\theta_1$, $\sigma$, $\chi$ and $2\chi$. The gray squares indicate $\sigma=1$,  and  the white squares indicate $\sigma=-1$, and the arrows indicating XY spins.

We observe only domain walls inside the $B_2$ phase and  the half-integer vortex is observed near the transition point between the $B_2$ and $C$ phases.
In Fig~\ref{fig:jb=0snapshot} (a), $\theta_1$ (red),  and $\sigma_i$ are shown.
$\theta_2$ is not shown for clarity purposes, but it can be obtained by Eq.~\ref{eq:map}.
Along different directions of the domain wall, the angle $\theta_1$ occurs a $\pi$-flip along the $\mathbb{Z}_2$ domain wall in the $y$-direction, while there is no domain wall  along the $x$-direction. Conversely, $\theta_2$ has a $\pi$-flip along the domain wall in the $x$-direction, while there is no in the $y$ direction (not shown).

Figure~\ref{fig:jb=0snapshot} (b) shows the configuration of $\chi=\theta_1 + \theta_2$.
It can be seen that the $\chi$ angle inside each shaded block is opposite to the surrounding neighboring spins. There are domain walls in both the $x$ and $y$ directions. Therefore, $M_{\chi}=M_{\theta_1+\theta_2}\approx0$ is disordered.
Although the actual value in the figure is finite
at $M_{\chi}\approx 0.079$, it is found to be zero in the subsequent finite size scaling.
In Fig.~\ref{fig:jb=0snapshot} (c) the configuration of $2\chi=2\theta_1 + 2\theta_2$ is shown.
For $2\chi$, the domain wall hardly appears anymore, and all the spins point approximately the same way, i.e., $2\chi$ is ordered. In the configuration, the actual value is $M_{2\chi}\approx 0.998$.

{Let's mathematically illustrate the $2\chi$ order, that is,  the equality of  $2\chi_{i,j}$ and $2\chi_{i+\vec{x},j}$  for two neighbouring sites.
From the configuration depicted in Fig.~\ref{fig:jb=0snapshot}, when the Ising variables satisfy $\sigma_{i,j}\sigma_{i+\vec{x},j} = 1 (-1)$, the XY spin variables then satisfy $\Delta\theta_{i,i+\vec{x}} = 0 (\pi)$.
Due to the strong interaction $\gamma$ in Eq.~\ref{eq:ham}, the  XY spins $\theta_{i,j,1}$ and $\theta_{i,j,2}$  between the upper and lower layers are perpendicular to each other,  i.e., $\theta_{i,j,2} = \theta_{i,j,1} + \sigma_{i,j}\frac{\pi}{2}$.
Therefore, the difference between  $2\chi_{i,j}$ and $2\chi_{i+\vec{x},j}$ should be}
{
\begin{align}
\Delta 2\chi_{i,i+\vec{x}}&=2(\chi_{i,j}-\chi_{i+\vec{x},j})\nonumber\\
&=2\left[\theta_{i,j,1}+\theta_{i,j,1}+\sigma_{i,j}\frac{\pi}{2}-\theta_{i+\vec{x},j,1}-\theta_{i+\vec{x},j,1}-\sigma_{i+\vec{x},j}\frac{\pi}{2}\right]\nonumber\\
&=4(\theta_{i,j,1}-\theta_{i+\vec{x},j,1})+\pi(\sigma_{i,j}-\sigma_{i+\vec{x},j})\nonumber\\
&=4\Delta\theta_{i,i+\vec{x}}+\pi(\sigma_{i,j}-\sigma_{i+\vec{x},j}).
\end{align}}
{
When $\Delta\theta_{i,i+\vec{x}}=0$ and $\sigma_{i,j}=\sigma_{i+\vec{x},j}$, it leads to   $\Delta 2\chi_{i,i+\vec{x}}=0$.  Similarly, $\Delta\theta_{i,i+\vec{x}}=\pi$ and $\sigma_{i,j}=-\sigma_{i+\vec{x},j}$,  it follows that $\Delta 2\chi_{i,i+\vec{x}}=4\pi\pm2\pi\Rightarrow 0$. The same derivation can also be applied in the $y$-direction. Therefore, $B_2$ phase has 2XY-order.}

\begin{figure}[t]
    \centering
    \includegraphics[scale=0.4]{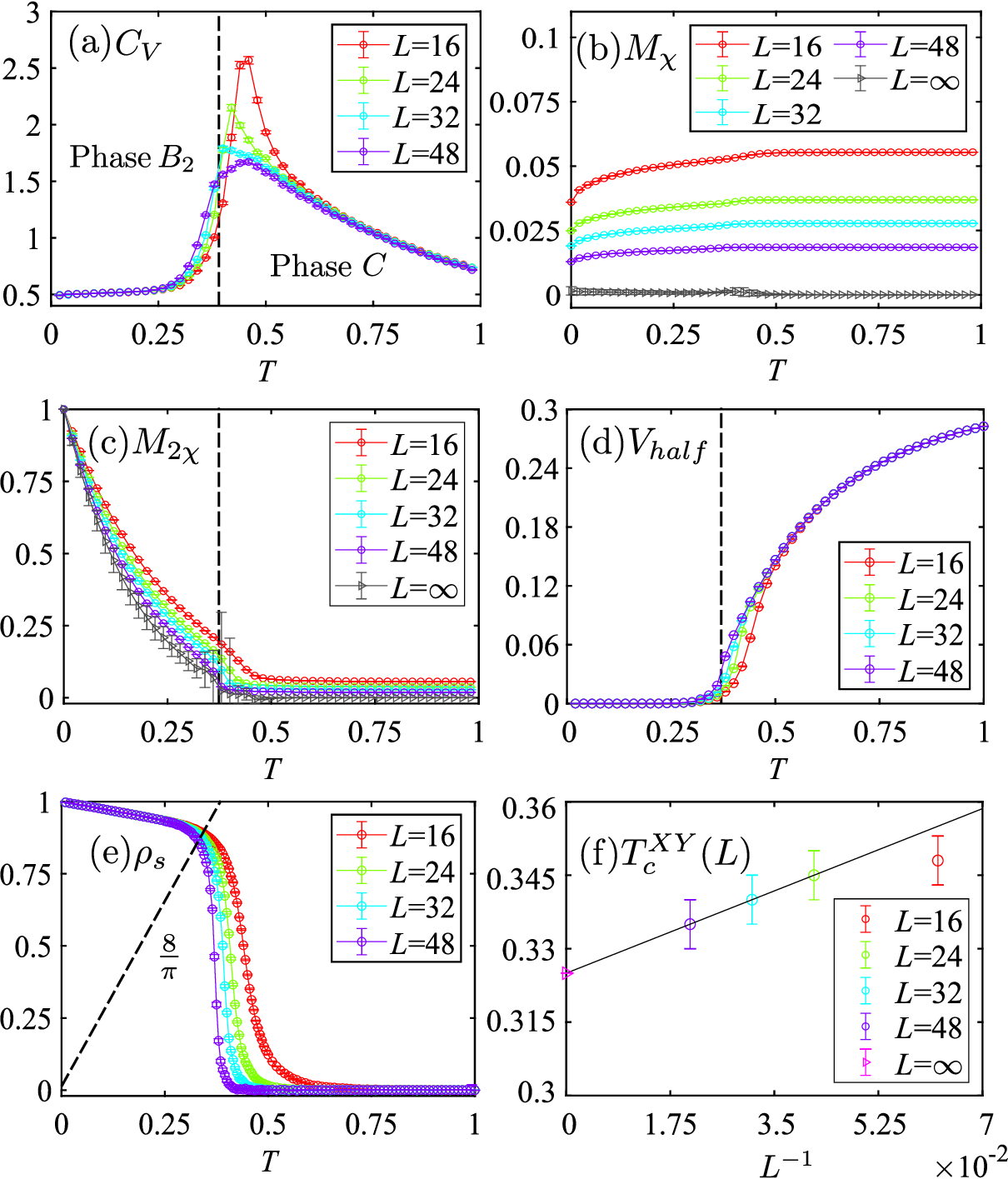}
    \caption{The quantities  for the XY variables at $J_a=1, J_b=0$, and $T_c^{XY}\approx 0.3250(1)$. (a) Specific heat $C_{V}$; (b) Magnetization $M_{\chi}$; (c) Magnetization $M_{2\chi}$; (d) the density of half-integer vortex (e) the spin stiffness $\rho_{s}$ (f) extrapolation to obtain the phase-transition point.}
    \label{fig:jb=0xy}
\end{figure}

Thus, the existence of the $B_2$ phase is confirmed and the problem becomes how to determine the phase transition point.
Firstly, we try to understand the phase transition  by the excitation of the topology. For simplicity, the picture with only one domain wall and one excited vortex pair is discussed in terms of free energy. Mean while, supposing  the position of domain wall is fixed, and therefore the entropy should be $\ln{1}=0$. For one domain wall, the free energy is 
\begin{equation}
    F_{DW}=E-TS\simeq J_{b}L_d-T\ln{1}=J_{b}L_d,
\end{equation}
{where  $L_d$ is the length of domain-wall, such as the blue line in Fig.~\ref{fig:jb=0snapshot} (b). 
    With the increasing of the temperature,
    a pair of half-vortex and anti-half-vortex emerges, and the new free energy becomes~\cite{cenkexu}}
\begin{align}
    F & = F_{DW} +F_{vortex-~pair}                                \\
      & =  J_{b}L_d+C_1\text{ln} L_d -T\text{ln} L_d\nonumber,
    \label{eq:ftot}
\end{align}
where $\text{ln} L_d$ and $C_1\text{ln} L_d$ are the  entropy and energy  of the vortex pair~\cite{Kardar_2007, PhysRevLett.55.541}, respectively. By solving for $F\le F_{DW}$, the rough phase transition temperature $T$ can be obtained.

{More accurate phase transition points still need to be determined by physical quantities calculated by numerical methods.}
Figure~\ref{fig:jb=0xy} (a) shows the specific heat $C_V$ in the range $T/J_a\in [0, 1]$ while $J_b=0$. Using data for sizes $L = 16-48$ it has been shown that, the non-divergent behavior of the specific heat peaks indicates the presence of the BKT phase transition.

As predicted, in both the $B_2$ and $C$  phases, $\chi$ is disordered ($M_{\chi}=0, L\rightarrow\infty$)  as shown in Fig.~\ref{fig:jb=0xy}(b).
However, obtaining such data in agreement with expectations is quite challenging.
First, the phase transition point corresponds to a very low temperature, around $T_c\approx0.34$.
Traditional cluster methods such as the Wolff algorithm are inefficient in this mixed system.
As introduced in Sec~\ref{sec:method}~~~~,
a combination of various methods including the Metropolis, Wolff, line-shaped cluster and PT methods, is used to obtain results consistent with expectations.

In order to obtain the phase transition point, we analyze   the signal of $2\chi$. The first quantity of interest is $M_{2\chi}$ as shown in Fig.~\ref{fig:jb=0xy} (c).
In the low-temperature phase with $T<T_c$,  $M_{2\chi}>0$  and continuously changes smoothly to 0.
This suggests that when we magnify the angle by a factor of two, the $2\chi$ ordered state is found from the $\chi$ disordered configuration and remains stable in the thermodynamic limit.

Additionally, we assess the density of half-integer vortices, as shown in Fig.~\ref{fig:jb=0xy} (d).
In the low-temperature phase, the density of half-integer vortices is zero.
Near the critical point, $T\approx T_c^{XY}$, half-vortex excitation is observed. This consists with the observations from the pure XY model~\cite{Hsieh_2013}, where vortex excitation is typically observed at temperatures $T\approx T_c^{XY}$.
In the high-temperature phase, numerous half-vortices become excited, resulting in a non-zero density.
Experimentally, the vortex distriubution  can also be observed~\cite{ex_vortex2,Klaus2022}.

\begin{figure}[t]
    \centering
    \includegraphics[scale=0.4]{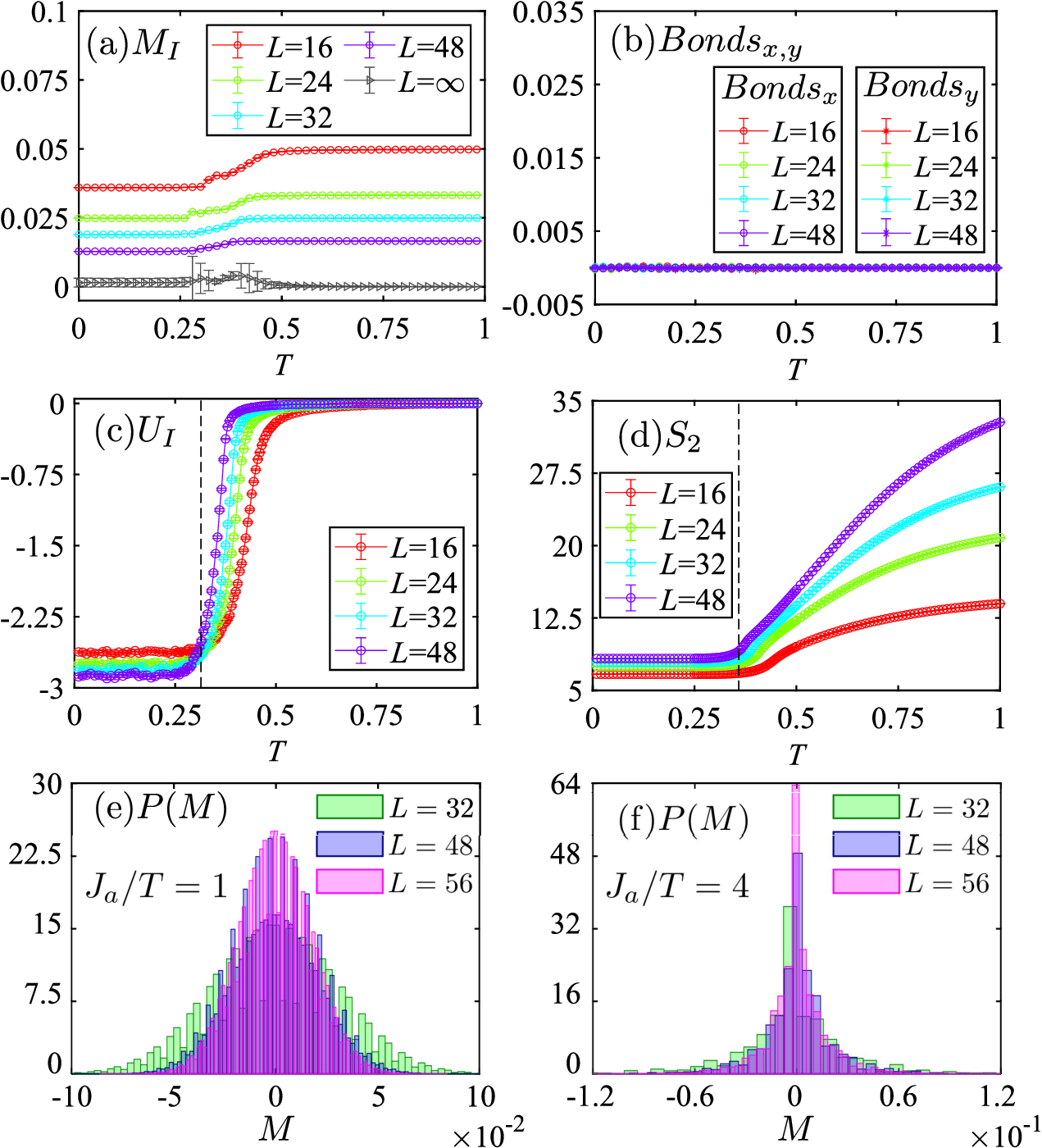}
    \caption{The quantities for the Ising variables ($J_a=1, J_b=0$).  (a) Magnetization $M_{I}$; (b) the bonds order $Bonds_{x,y}$; (c) Binder cumulant $U_{I}$; and (d) the percolation susceptibility $S_{2}$.
        (e) Distribution of $M_{I}$ in the $C$ phase; (f) the distribution of $M_{I}$ in the $B_{2}$ phase.}
    \label{fig:jb=0ising}
\end{figure}

In addition, inspired by the formula for the superfluid density of isotropic systems given in the literature~\cite{sandvik}, we derive an expression for the spin stiffness in the anisotropic case as shown from Eq.~\ref{eq:sf} to Eq.~\ref{eq:sfi}.
Within the BKT theory, spin stiffness exhibits a universal jump at $T_c$~\cite{wessel_xy},
\begin{equation}
    \rho_s = \frac{2}{\nu^2 \pi} T_c, ~~\nu = 1/2.
\end{equation}
The line $8T/\pi$, the spin stiffness data $\rho_s$ of various sizes in Fig.~\ref{fig:jb=0xy}(e) intersect at the location of the jump $T_c\approx 0.325(5)$ and the fitting $T_c^{XY}(L)$ vs $L^{-1}$ is shown in Fig.~\ref{fig:jb=0xy}(f).

Next, the phase transition is analyzed from the perspective of the Ising variables. Due to the numerous domain walls in both the $B_2$ and $C$ phases, the Ising variables are disordered.
Clearly, in Figs.~\ref{fig:jb=0ising}(a) and (b), $M_I$ is zero  as $L\rightarrow \infty$. The average bonds in both directions,  described in Eq.~\ref{eq:bond}, are also zeros.
However, we see the phase transition signal in terms of the Binder ratio $U_I$ corresponding to the Ising variable as well as the percolation susceptibility $S_2$ in Figs.~\ref{fig:jb=0ising}(c) and (d).
In the  $C$ phase, the shape of $P(M)$  should be Gaussian as shown in Fig.~\ref{fig:jb=0ising}(e).
Due to the relations
\begin{equation}
    \int_{-\infty}^{\infty} M^2 \frac{1}{\sqrt{2\pi}} e^{-\frac{M^2}{2}} dM =1,
\end{equation}
and
\begin{equation}
    \int_{-\infty}^{\infty} M^4 \frac{1}{\sqrt{2\pi}} e^{-\frac{M^2}{2}} dM=3,
\end{equation}
we can get $ U_{I}=\frac{3}{2}\left(1-\frac{1}{3}\expval{M^{4}}/\expval{M^2}^2\right)\rightarrow 0$~\cite{binder}. In the $B_2$ phase, characterized by the quasi-exponential distribution  $P(M) =\lambda e^{-\lambda |M|}$ as shown in Fig.~\ref{fig:jb=0ising} (f),  we can get $U_{I}<0$. The change in $U_I$ from 0 to less than zero reflects the phenomenon of phase transition.
Usually, a negative peak of $U_I$ represents a first-order phase transition~\cite{binder}.
A negative $U_{I}$, without a negative peak, is not indicative of a first-order phase transition.


\subsection{Signatures of our proposed \texorpdfstring{$M_\theta$}{}}
\label{sec:mtheta}
\begin{figure}[t]
    \centering \includegraphics[scale=0.4]{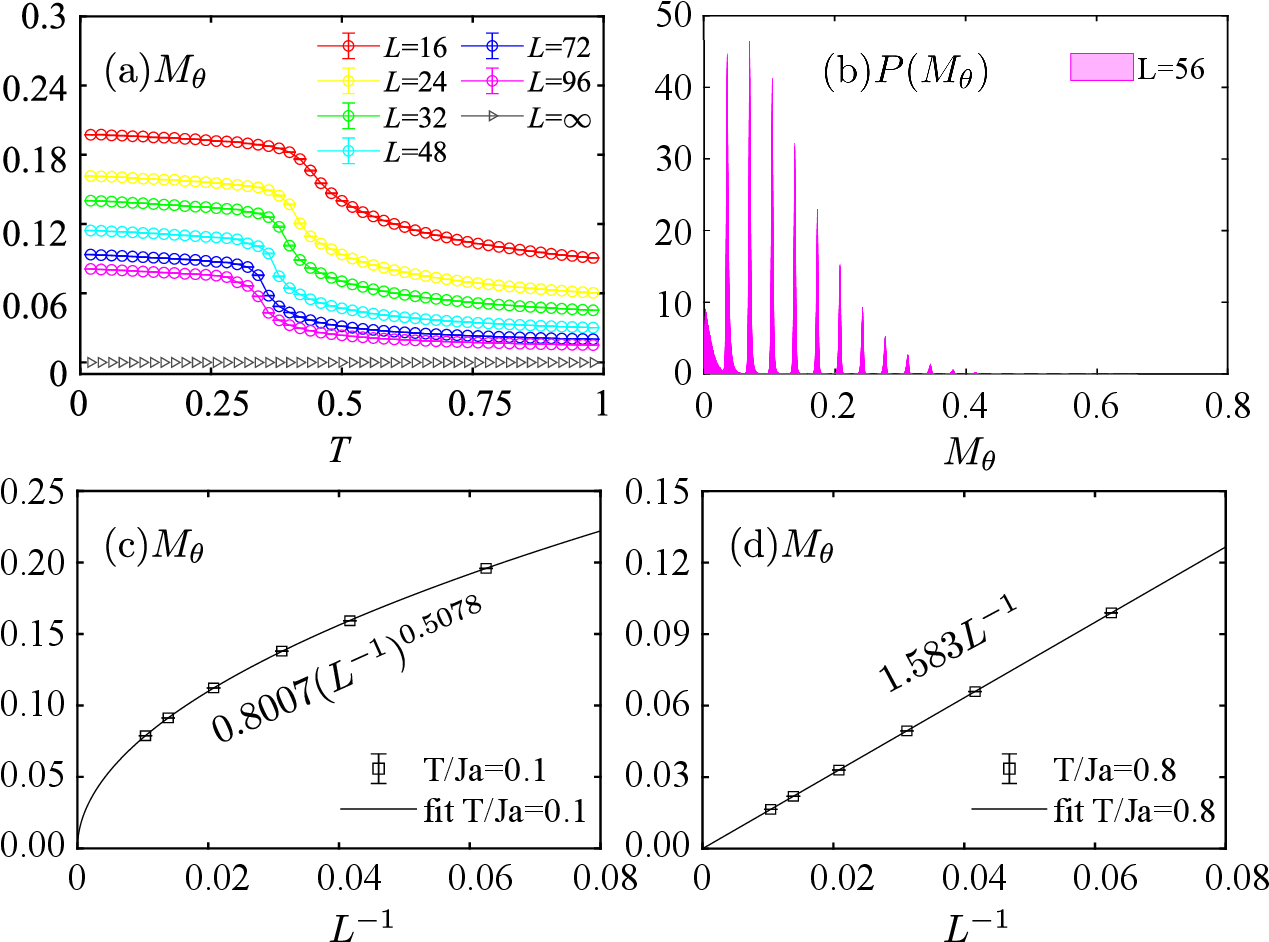}
    \caption{ \sethlcolor{green}{The signatures where ($M_{\theta} = 0$) occur in the thermodynamic limit. }(a) $M_{\theta}$ vs $T$; (b) the probability distributions of $M_{\theta}$ for sizes $L = 56$ when $T/J_a=0.1$; (c)  and (d) show $M_{\theta}$ vs $L^{-1}$ at different \sethlcolor{green}{phases}.}
    \label{label:app1}
\end{figure}

In this subsection,  we list  the behaviours of physical quantity $M_{\theta}$, in different phases, as shown in Fig.~\ref{label:app1}.
Because the variables for our numerical simulation are XY spins on one layer $\theta$  and Ising variable $\sigma_i$,
one might ask whether or not the magnetisation  $M_{\theta}$ is 0, except that $M_{\chi}$ is zero in phase $B_2$.

\begin{figure}[t]
    \centering
    \includegraphics[scale=0.4]{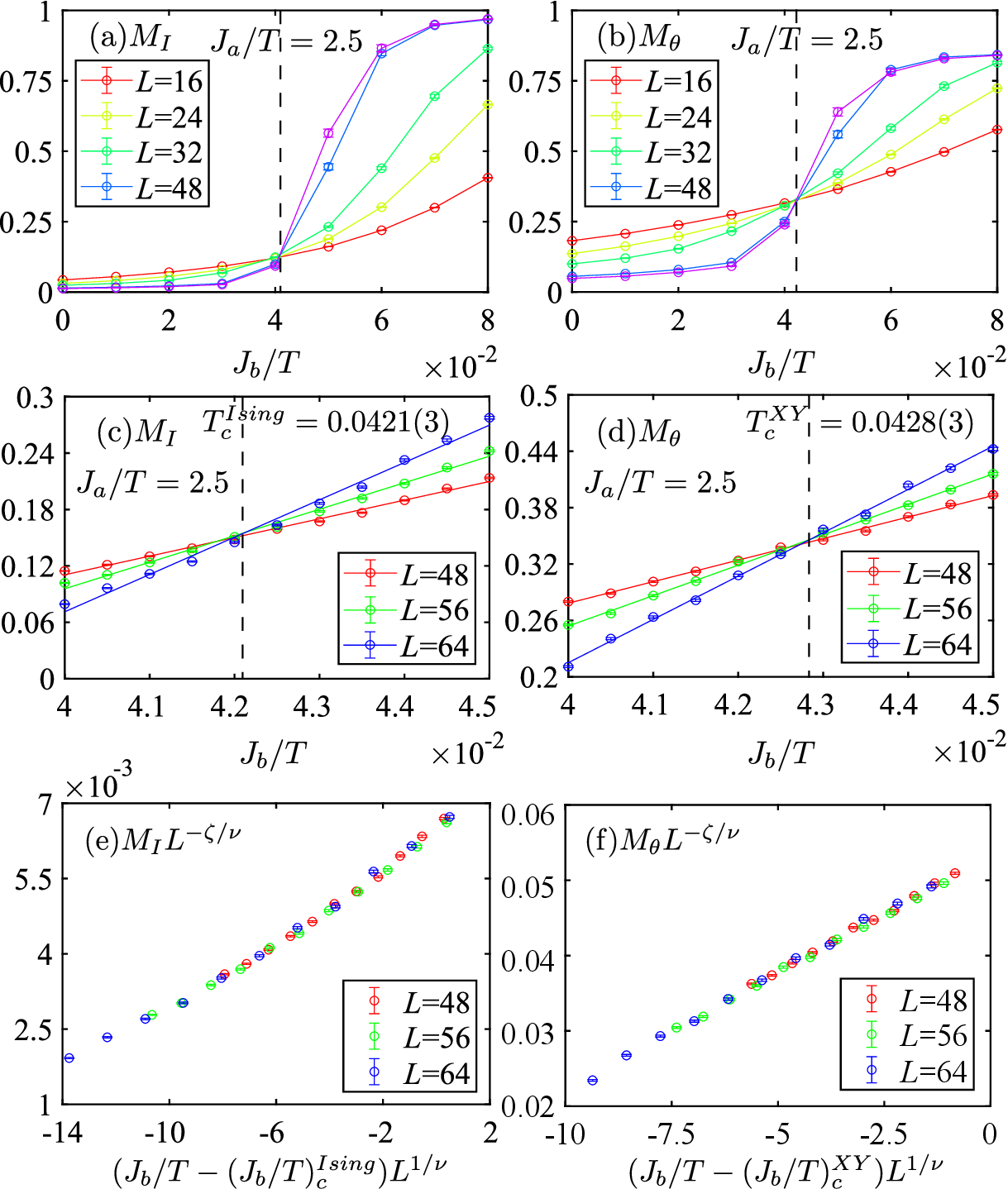}
    \caption{ (a)-(d)
        $M_I$ and $M_{\theta}$ vs $J_b/T$ at $J_a/T=2.5$. The intersection positions are close to 0.0421(3) and 0.0428(3), respectively, indicating that the two phase transitions are not at the same location.  (e)-(f) \sethlcolor{green}{The critical exponent  $\nu$ obtained by finite size scaling is close to $0.5$, which characterize the  first order phase transition.} }
    \label{fig:vert1}
\end{figure}

In Fig.~\ref{label:app1} (a), $M_{\theta}$ vs $T$ is plotted. In the thermodynamic limit, it is not difficult to understand that $M_{\theta}$ is zero in the disordered $C$ phase.\sethlcolor{green}{
    The $B_2$ phase has no $\chi$-order with $M_{\chi}=0$.  Additionally, it exhibits no $\theta$-order with $M_{\theta}=0$ .}

Next, we analyze $M_{\theta}$ through the probability distribution $P(M_{\theta})$ as shown in Fig.~\ref{label:app1} (b) at $J_b/T=0$.
The distribution of $M_{\theta}$ has many discrete peaks.
The horizontal coordinates corresponding to these peaks can actually be calculated manually.
As shown by the red arrows in the snapshot in Fig.~\ref {fig:jb=0snapshot} (a), these arrows are almost identical in the vertical direction, i.e. the stripe pattern. Therefore, we simply pick one of the rows of spins to analyze.

The first peak corresponds to a horizontal coordinate of 0, which represents the configuration of the system where there are $L/2$ up spins and $L/2$ down spins.
In this case, the probability $P(M_{\theta}=0)$ is proportional to the  coefficient $C_{L}^{L/2}$.
For the second peak, there are $L/2+1$ up spins and $L/2-1$ down spins. Taking $L=56$ for example, the location  is  $(29-27)/56$ and its probability proportional to $2C_{L}^{L/2-1}$.
Thus the ratio of the areas of the two peaks is $C_{L}^{L/2}:2C_{L}^{L/2-1}=0.5208$, which coincides with the results of the MC simulation. We also examine the results at other sizes and other peaks and the conclusions remain the same. The peaks at different positions can precisely reflect the excellent ergodic capability of our method.

\sethlcolor{green}{In fact, these highly degenerate states are caused by domain wall excitations without cost energies.  By analyzing the first line in Eq.~\ref{eq:ixy}, In the $x$ direction,  for the two most neighboring spins, there are two possible states. one  state is  $\sigma_{i,j}\sigma_{i+x,j}=-1$ and $\theta_{i+x,j}=\theta_{i,j}+\pi$.
    and  Each term involved in the summation in the first line of Eq.~\ref{eq:ham} becomes}
\begin{align}
    E_x & = -\left(J_b +\underline{ J_a \sigma_{i,j} \sigma_{i+\vec{x},j}}\right) \cos(\theta_{i,j} - \theta_{i+\vec{x},j}) \nonumber \\
        & =  -(J_b\underline{-J_a})\cos(-\pi)=J_b-J_a.
    \label{eq:ftot}
\end{align}
\sethlcolor{green}{The other state is  $\sigma_{i,j}\sigma_{i+x,j}=1$ and $\theta_{i+x,j}=\theta_{i,j}$.
    and Each term involved in the summation in the first line of Eq.~\ref{eq:ham} becomes}

\begin{align}
    E_x^{'} & = -\left(J_b +\underline{ J_a \sigma_{i,j} \sigma_{i+\vec{x},j}}\right) \cos(\theta_{i,j} - \theta_{i+\vec{x},j})\nonumber \\
            & =  -(J_b\underline{+J_a})\cos(0)=-J_b-J_a,
    \label{eq:ftot}
\end{align}
\sethlcolor{green}{ For case the $J_a\gg J_b=0$ or  $J_a\gg J_b>0$, $E_x$ is equal or very close to  $E_x^{'}$. Thus, any combinations of the number of $E_x$ and $E_x'$ counterparts are energy-equivalent. The analysis of the interaction in the $y$-direction is similar.
Totally speaking, when domain walls appear, the configurations don't have Ising-order or XY-order but instead have 2XY-order.}

In both the $B_2$ and $C$ phases, as the system size approaches infinity in the thermodynamic limit, $M_{\theta}$ converges  to 0, but it varies with size in different ways.
In Fig.~\ref{label:app1} (c), the plot shows
$M_{\theta}$ vs. $L^{-1}$  at $T/J_a=0.1$ in the $B_2$ phase. The data are well fitted with the function  $0.8007L^{-0.5078}$.
In Fig.~\ref{label:app1} (d), in the $C$ phase, the plot of $M_{\theta}$ vs. $L^{-1}$  is well fitted with the function  $1.583L^{-1}$ and it decays faster.

\sethlcolor{green}{In short, the phase $B_2$ and phase C, although both have no $\theta$-order, have distinctly different patterns.}
In the $B_2$ phase, the pattern for $\theta$ (red arrows) appears striped, showing disorder only in the $x$  direction. But in the $C$ phase, the spin is completely disordered, lacking any noticeable pattern.
\begin{table}[thb]
    \centering
    \caption{List of parameters and dimensions corresponding to the double-peaked distribution in Fig. \ref{fig:Ja=1.5-2.7}.}
    \label{tab:Ja=1.5-2.7_para_table}
    \begin{tabular}{ccccc}
        $Ja/T$ & $L$ & $J_{b}/T(P(E))$ & $J_{b}/T(P(M_{I}))$ & $J_{b}/T(P(M_{\theta}))$ \\
        \midrule
        1.5    & 48  & 0.2600          & 0.2585              & 0.2580                   \\
        1.5    & 72  & 0.2620          & 0.2610              & 0.2606                   \\
        \hline
        1.7687 & 48  & 0.1761          & 0.1761              & 0.1761                   \\
        1.7687 & 72  & 0.1766          & 0.1766              & 0.1766                   \\
        \hline
        2.0    & 48  & 0.1188          & 0.1188              & 0.1184                   \\
        2.0    & 72  & 0.1194          & 0.1194              & 0.1192                   \\
        \hline
        2.7    & 48  & 0.0360          & 0.0440              & 0.0370                   \\
        2.7    & 72  & 0.0340          & 0.0360              & 0.0320                   \\
        \bottomrule
    \end{tabular}
\end{table}

\begin{figure*}[p]
    \centering
    \includegraphics[width=1\linewidth]{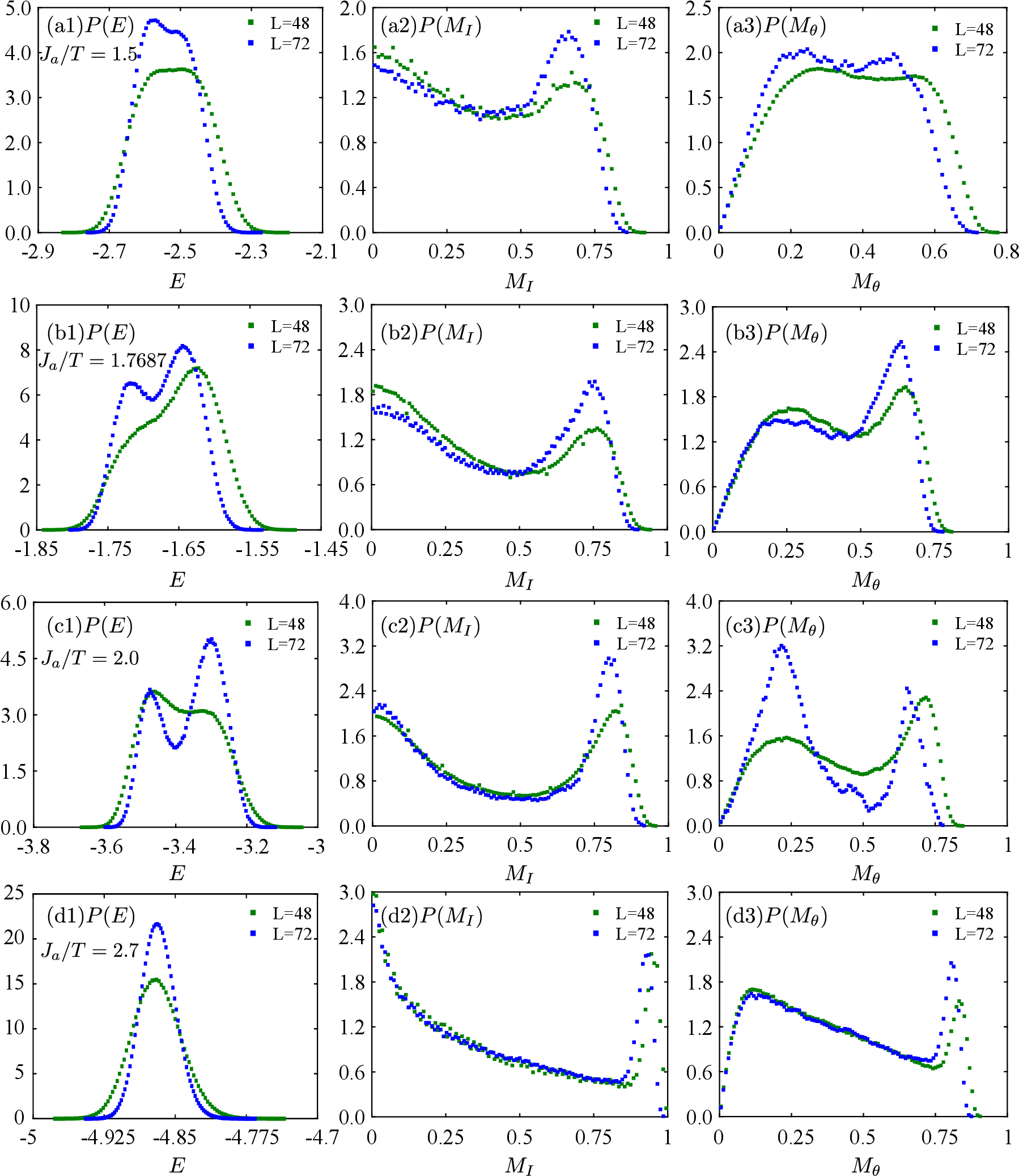}
    \caption{The probability distribution (normalized) of physical quantities near the critical point obtained by fixing the parameter $J_{a}/T$ and scanning $J_{b}/T$. This two-peaked distribution characterizes the first-order phase transition. Details of the  parameters are shown in Table \ref{tab:Ja=1.5-2.7_para_table}.}
    \label{fig:Ja=1.5-2.7}
\end{figure*}

\subsection{ First order transitions along \texorpdfstring{$J_a/T=1.5- 2.7$}{}}
\label{sec:1st}
In the global phase diagram in Fig.~\ref{fig:global}, the regimes marked in green represent first-order phase transitions.
Two questions need to be addressed. The first question is whether the Ising first-order phase transition and the XY first-order phase transition occur with the same parameter? Second, what is the signal for the first-order phase transition?

In the results shown in Fig.~\ref{fig:vert1}, it is observed that in the region of the first-order phase transition, the phase transitions for the Ising and XY variables still do not occur simultaneously.
Fixing $J_a/T=2.5$, $M_I$ and $M_{\theta}$ vs. $J_b/T$ are plotted in the range $0-0.08$ and  the zoomed range $0.04-0.045$.
Apparently, the intersection of $M_I$ lines of different sizes is close to 0.0421(3) while the intersection of $M_{\theta}$ lines of different sizes is close to 0.0428(3).
Therefore, the two transitions do not occur simultaneously. In this parameter interval, the theoretical predictions~\cite{cenkexu} of the phase diagram structure and the numerical experiments are in agreement.

In response to the second question, when testing the Binder ratio $U_I$ for different sizes, no intersection point is found, but a negative peak is observed~\cite{binder} (although not shown).
In Figs.~\ref{fig:vert1} (a)-(d), it is intriguing to observe that the magnetization curves, represented by $M_I$  and $M_{\theta}$, for different system sizes intersect at the critical points. These intersections are attributed to the symmetry breaking of continuous variables~\cite{kbinder}.
{
    In Figs.~\ref{fig:vert1} (e) and (f), the critical exponents are obtained by using finite-size scaling~\cite{melchert2009autoscalepy}: $\nu \approx 0.52(1)$ for the Ising phase transition and $\nu \approx 0.5(2)$ for the XY phase transition.
    The values of the correlation length exponent $\nu$ are either equal to or close to the theoretical value of $1/d=0.5$ for the first-order phase transition~\cite{034006, kbinder}. More precisely, the value $0.52(1)$ is greater than $0.5$ due to the effect of the so-called weak first-order transitions~\cite{034006}. }

Furthermore, the distributions $P(E)$, $P(M_I)$, $P(M_\theta)$ are plotted to check whether or not double peaks emerge. At the beginning of the green line, with parameter  $J_a/T=1.5$, the signatures of the double peaks are weak or even absent, indicating a phenomenon known as pseudo first-order phase transition~\cite{sandvik_1stph,ATPFO}. As the system size increases, the two peaks in the energy distribution approach each other, as illustrated in Fig.~\ref{fig:Ja=1.5-2.7} (a).

In the regime $J_b/T=0.1$, $J_a/T=2$ and $J_a/T=1.7867$  in Fig.~\ref{fig:global}, the signature of the first-order transition is  the most obvious. The evidence is that the double peaks emerge for  the distributions $P(M_I)$, $P(M_{\theta})$ and $P(E)$ with $L=48, 72$ as shown in Figs.~\ref{fig:Ja=1.5-2.7} (b) and (c). In the end of the green line $J_a/T=2.7$, the double peaks of energy disappear.
Although both  $P(M_I)$ and $P(M_{\theta})$ exhibit bimodal distributions, which indicates energy degeneracy in the system or coexistence of two phases in the system, it is still not possible to determine whether a first-order phase transition occurs at this point.

It is worth noting that the double peaks observed in
$M_I$ and $M_{\theta}$ do not occur at the same parameter values of  $J_a/T$ and $J_b/T$. There is a deviation between the parameters corresponding to $M_I$ and $M_{\theta}$ for different system sizes.
The specific values of the parameters are listed in the Table~\ref{tab:Ja=1.5-2.7_para_table}. The parameter difference for $M_I$ and $M_{\theta}$, obtained from the same lattice size, indicates that the two phase transitions, do not occur simultaneously.
\begin{figure}[tbh]
    \centering
    \includegraphics[scale=0.4]{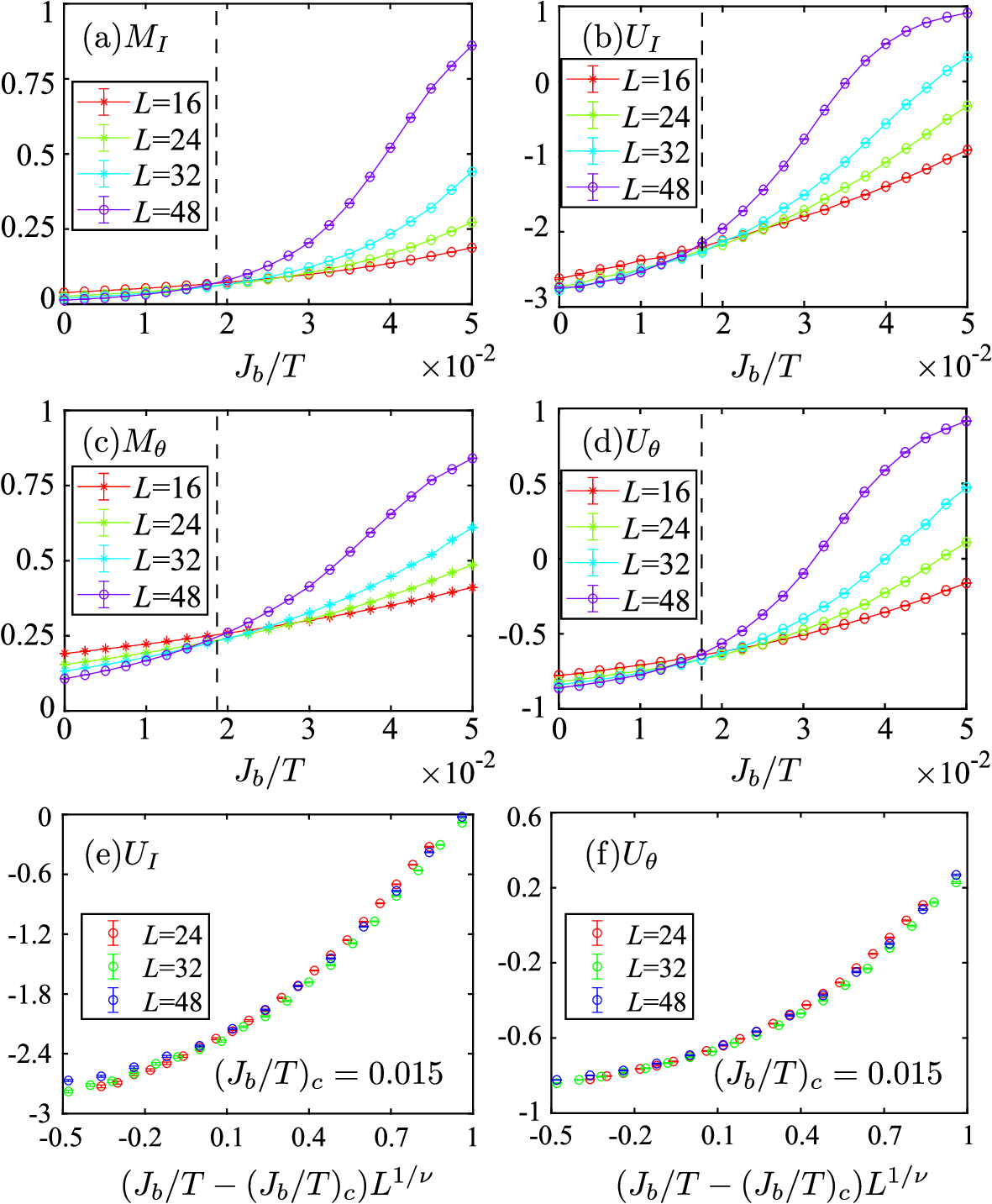}
    \caption{$J_{a}/T=3.5$, (a)$M_{I}$ vs $J_{b}/T$; (b)$U_{I}$ vs $J_{b}/T$; (c)$M_{\theta}$ vs $J_{b}/T$; (d)$U_{\theta}$ vs $J_{b}/T$, \sethlcolor{green}{(e) $U_I $ vs $(J_b/T-(J_b/T)_c)L^{1/\nu}$ (f)$U_{\theta} $ vs $(J_b/T-(J_b/T)_c)L^{1/\nu}$, where $\nu\approx 0.5$. }}
    \label{fig:B2A}
\end{figure}

\subsection{\texorpdfstring{$J_a/T$=3.5, transitions between the $A$ and $B_2$ phases}{} }
\label{sec:Ja=3.5}

In Fig.~\ref{fig:B2A}, we explore the phase transition  between the A-$B_2$ phase.
By fixing $J_a/T=3.5$ and scanning  $J_b/T$, at this point, we observe that both $M_I$ and $M_{\theta}$ for different sizes exhibit common intersections,  as reported in Ref.~\cite{kbinder} recently.

The Binder cumulants $U_I$ and $U_{\theta}$ exhibit similar behavior during the phase transition. By performing a data collapse analysis for these two quantities, we plot $U_{I}$ and $U_{\theta}$ as functions of $(T-T_c)L^{1/\nu}$, for different system sizes. Remarkably, the data points for various sizes overlap with each other. This observation indicates that the critical exponent of the phase transition follows the Ising universal class with $\nu=1$.

\section{Conclusion}
\label{sec:con}
In this paper, we investigate the spatially anisotropic Ising-XY models using our proposed line-shaped cluster-updating  MC scheme and other methods.
Our method allows for spin flips in clusters without rejection in the anisotropic limit when $J_b = 0$, following the energy invariance principle. Even at small but nonzero values of $J_b$, we can still perform cluster-spin flips using the Metropolis algorithm. Our approach effectively addresses the non-ergodic challenges (see Fig.~\ref{fig:compare}(b)) faced by traditional Monte Carlo methods in the anisotropic limit with significant degeneracy and low temperatures.
The main findings of this study are as follows:

(I) The nature of the $B_2$  phase has been investigated in depth.
In the anisotropic limit, the  $B_2$ phase is  Ising and XY disorder,  but 2XY order, meaning that the magnetizations have the properties $M_I=0$, $M_{\chi}=0$,  and  $M_{2\chi}\ne 0$.
Furthermore, the distribution $P(M_\theta)$ even can be analyzed manually due to the XY spins exhibit a quasi-one-dimensional pattern.
Our proposed $M_\theta(L)$ in the $B_2$ phase follows a scaling behavior of  $M_\theta(L)=0.8007L^{-0.5078}$, while in the high temperature disorder $C$ phase, it follows a different scaling behavior of $M_{\theta}(L)=1.583L^{-1}$.
We demonstrate that the superfluid density $\rho_s$ scales as $\rho_s = \frac{2}{\nu^2 \pi} T_c$, with the critical exponent  $\nu = 1/2$.

(II) In the constructed  global phase-diagram, the two boundary lines are very close to each other and differ significantly from the schematic phase diagram reported earlier.
In the isotropic limit, the critical points of the Ising are XY phase transitions have been identified  at $T_c^I=1.361(1)$ and $T_c^{XY}=1.342(2)$, respectively.
In the anisotropic limit, for example $J_a/T=2.5$, the critical points locate at $J_b/T_c^{Ising}=0.0421(3)$ and $J_b/T_c^{xy}=0.0428(3)$.

(III) In the parameter regions that deviate from the isotropic  limits, both phase transitions become first-order phase transitions.
Interestingly, $M_I$ and $M_{\theta}$ of different sizes intersect at a point due to continuous symmetry breaking~\cite{kbinder}. These phenomena are determined by the double-peaked distribution of the energy $E$ and magnetization $M_I$ and $M_{\theta}$.  Also, the critical exponent $\nu$  approximately equal to the inverse of the space dimension, i.e., $\nu=1/d$.

Despite conducting numerous numerical simulations, several open questions persist in this research. First, it remains uncertain whether there are novel critical exponents, particularly in the proximity of the first-order transition, as discussed in~\cite{isxy2st}. Second, the applicability of the coupled Ising-XY model extends beyond two-dimensional lattice systems, potentially revealing novel phenomena in three-dimensional systems, as explored in Ref.~\cite{cenkexu}.
Furthermore, the direct investigation of the Bose-Hubbard model with $p_x$ and $p_y$ orbitals holds promise for uncovering half-vortices~\cite{cenkexu}. Similar Hamiltonians have already been simulated directly, as referenced in~\cite{2bh}. On the experimental front, the realization of a two-component Bose-Hubbard system can be achieved using $^{87}\text{Rb}$ cold atoms in two different hyperfine states~\cite{prx,Meng2023}, or through an array of cavity polaritons~\cite{pra}. These experimental avenues provide exciting opportunities for further exploration in this field.


\vskip 0.5 cm
\addcontentsline{toc}{chapter}{Appendix A: Appendix section heading}
\section*{\textcolor{black}{Appendix A:  Direct simulation of Eq.~\ref{eq:ham} for the bilayer XY model}}
{
    Eq.~\ref{eq:ixy} comes from the mapping of Eq.~\ref{eq:ham}. In our simulation, we directly use Eq.~\ref{eq:ham} and compare it with the results from Eq.~\ref{eq:ixy}. This helps verify the accuracy of the mapping. Transforming an equation can change its underlying physics, so it's crucial to check if the angles between the two layers are perpendicular, i.e., if Eq.~\ref{eq:map} holds true or if
    $\Delta\theta_{i,j}=\theta_{i, j,1} -  \theta_{i,j, 2}=\pm \pi/2$.}

\begin{figure}[hbt]
    \centering
    \includegraphics[scale=0.4]{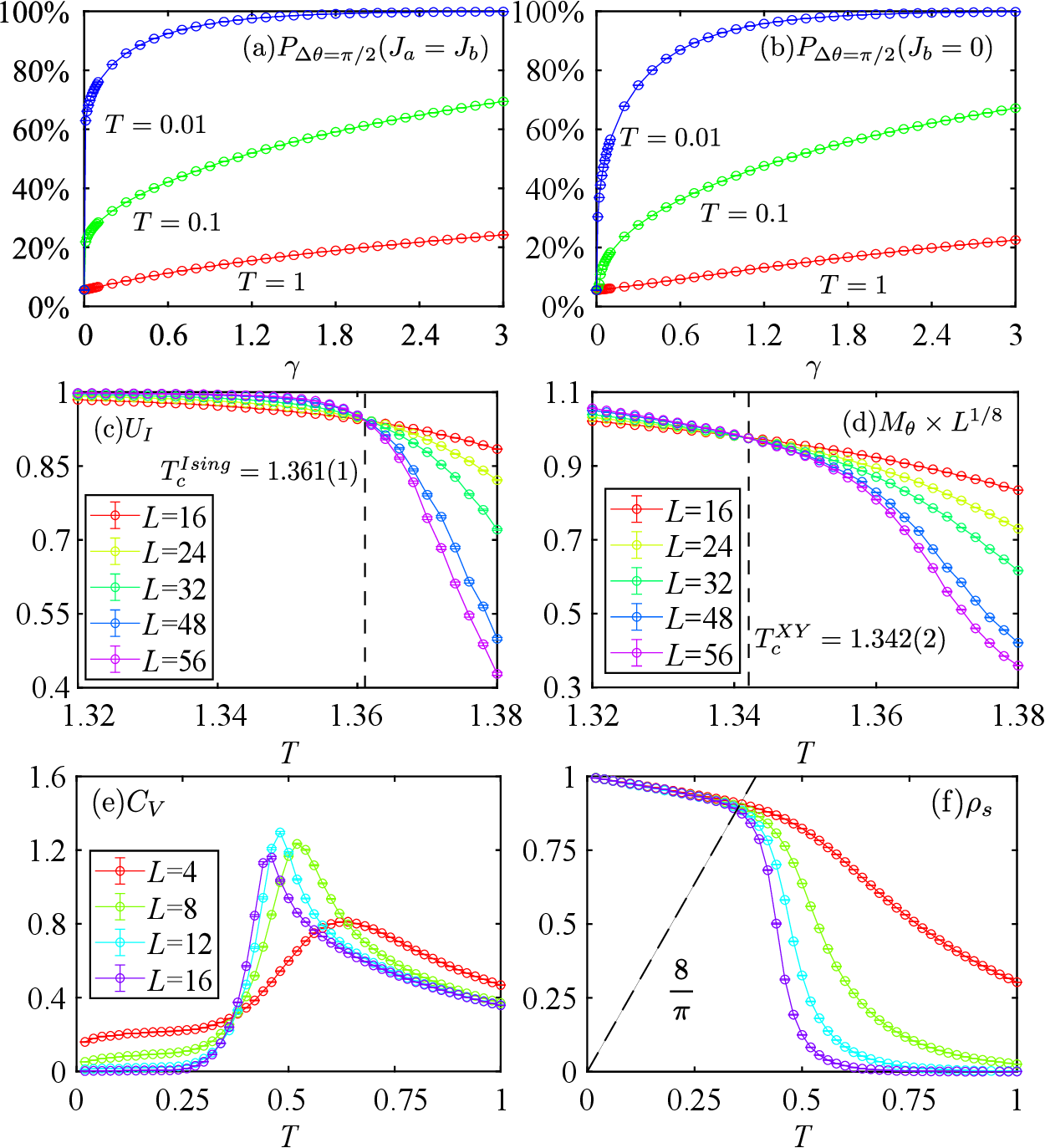}
    \caption{ 
    \textcolor{black}{ The results of direct simulating Eq.~\ref{eq:ham}. At different temperatures $T$=0.01, 0.1 and 1, the ratios of $\Delta\theta=\pm \pi/2$ are estimated as functions of $\gamma$ (a) with parameters $J_a=J_b=1$ and (b) with $J_b=0$, respectively.}
    The results of (c) $U_I$ and (d) $M_\theta$ with parameters $J_a=J_b$ and $\gamma=10^9$.
    (e) $C_V$ and (f) $\rho_s$ with $J_b=0$ and $\gamma=10^9$. $T_c=0.33(2)$ is obtained in (f). }
    \label{fig:15}
\end{figure}

 In Fig.~\ref{fig:15}(a) with parameters $J_a=J_b=1$ and 
(b) with $J_b=0$, the ratios of $\Delta\theta=\pm \pi/2$ are estimated as functions of $\gamma$, respectively. At low temperatures, specifically when  $\gamma>2$,
    the upper and lower neighboring spins are entirely perpendicular to each other.
    However, at a temperature of  $T=1$, larger values of  $\gamma$ are required, such as $\gamma=10^9$.    In Figs.~\ref{fig:15} (c)-(d), for $J_a=J_b=1$, using Eq.~\ref{eq:ham}, we recauculate the results for  $U_I$ and $M_\theta$, and obvserve that the critical points remain at
$T_c^{Ising}=1.361(1)$ and $T_c^{XY}=1.342(2)$, which are consistent with the results in Fig.~\ref{fig:A=B_phase_transition}.   In Figs.~\ref{fig:15} (e)-(f), with $J_b=0$, we recalculate the values of $C_V$ and $\rho_s$, and find that the critical point is still  at   $T_c^{XY}=0.33(2)$, consistent with the results $T_c^{XY}=0.3250(1)$ in Figs.~\ref{fig:jb=0xy} (e) and (f) within the error bars. 

\vskip 0.5 cm


\addcontentsline{toc}{chapter}{Acknowledgment}
\section*{Acknowledgment}
We would like to thank valuable discussion with Cenke Xu and Dingyun Yao, and the helpful suggestions from the anonymous reviewer.  This work was supported by the Hefei National Research Center for Physical Sciences at the Microscale (KF2021002), and project 12047503 supported by NSFC.
C.D. was supported by the National Science Foundation of China (NSFC) under Grant Numbers 11975024 and the Anhui Provincial Supporting Program for Excellent Young Talents in Colleges and Universities under Grant No. gxyqZD2019023.
Y.D. was supported by the National Natural Science Foundation of China under Grants No. 12275263, and the National Key R\&D Program of China (Grant No. 2018YFA0306501).

\begin{singlespace}
    \providecommand{\newblock}{}

\end{singlespace}
\end{document}